\documentclass[pre, amsmath, amssymb, showpacs,superscriptaddress, twocolumn]{revtex4-1}
\usepackage{bm}
\usepackage[mathcal]{eucal}
\usepackage{bbm}
\usepackage{nicefrac}
\usepackage{graphicx}
\usepackage{color}
\usepackage{microtype}

\renewcommand{\vec}[1]{\bm{#1}}
\newcommand{\uvec}[1]{\hat{\vec{#1}}}

\newcommand{\avr}[1]{\left\langle#1\right\rangle}
\newcommand{\bra}[1]{\left\langle#1\right|}
\newcommand{\ket}[1]{\left|#1\right\rangle}

\newcommand{\Lv}{\mathcal{L}}
\newcommand{\U}{\mathsf{U}}
\renewcommand{\P}{\mathcal{P}}
\newcommand{\Q}{\mathcal{Q}}

\newcommand{\LT}{\textsf{LT}}
\newcommand{\FT}{\textsf{FT}}

\DeclareMathOperator*{\argmin}{arg\,min}

\begin{document}

\title{The Glass Transition in Driven Granular Fluids: A
  Mode-Coupling Approach}
\author{W. T. Kranz} 
\affiliation{Georg-August-Universit\"at G\"ottingen,  Institut f\"ur 
  Theoretische Physik, Friedrich-Hund-Platz 1, 37077 G\"ottingen,
  Germany} 
\affiliation{Max-Planck-Institut f\"ur Dynamik und Selbstorganisation,
  Am Fa\ss berg 17, 37077 G\"ottingen, Germany}
\author{M. Sperl} \affiliation{Institut f\"ur Materialphysik im Weltraum,
Deutsches Zentrum f\"ur Luft- und Raumfahrt (DLR), 51170 K\"oln, Germany}
\author{A. Zippelius} 
\affiliation{Georg-August-Universit\"at G\"ottingen,  Institut f\"ur 
  Theoretische Physik, Friedrich-Hund-Platz 1, 37077 G\"ottingen,
  Germany} 
\affiliation{Max-Planck-Institut f\"ur Dynamik und Selbstorganisation,
  Am Fa\ss berg 17, 37077 G\"ottingen, Germany}


\date{\today}

\begin{abstract}
  We consider the stationary state of a fluid comprised of inelastic
  hard spheres or disks under the influence of a random,
  momentum-conserving external force. Starting from the microscopic
  description of the dynamics, we derive a nonlinear equation of
  motion for the coherent scattering function in two and three space
  dimensions. A glass transition is observed for all coefficients of
  restitution, $\varepsilon$, at a critical packing fraction,
  $\varphi_c(\varepsilon)$, below random close packing. The divergence
  of timescales at the glass-transition implies a dependence on
  compression rate upon further increase of the density - similar to
  the cooling rate dependence of a thermal glass. The critical
  dynamics for coherent motion as well as tagged particle dynamics is
  analyzed and shown to be non-universal with exponents depending on
  space dimension and degree of dissipation.
\end{abstract}

\maketitle

\section{Introduction}
\label{sec:introduction}

A wide range of fluids can be quenched into a disordered, solid
state. This includes metallic melts \cite{greer95}, colloidal
suspensions \cite{vanmegen95}, foams \cite{hoehler+cohen05} and
recently, evidence was given that granular fluids may also undergo a
glass transition
\cite{marty+dauchot05,abate+durian06,goldman+swinney06,reis+ingale07,keys+abate07,kranz+sperl10,sperl+kranz12}. 
Among all these different systems, colloidal suspensions are probably
best understood. Experiments by van Megen \textit{et al.}
\cite{pusey+vanmegen86,vanmegen95} showed that besides the fluid and
the ordered crystalline phase, colloidal suspensions in thermal
equilibrium can also form colloidal glasses: A dynamically arrested
state of the system which is characterized by diverging relaxation
times \cite{debenedetti+stillinger01}. While a complete theoretical
understanding of the glass transition in fragile glass formers is
still missing \cite{cavagna09}, mode coupling theories can quite
successfully describe some of the phenomena on a quantitative level
\cite{goetze09}.

One interesting question has been raised more recently: Does the glass
transition survive, if the system is driven by external forcing into a
nonequilibrium state? Or more generally, can one observe a glass
transition also in a nonequilibrium system? A well studied example are
sheared colloidal suspensions for which it was shown that the
equilibrium glass transition disappears
\cite{fuchs+cates02,fuchs+cates09}.  Another example is nonlinear
microrheology \cite{habdas04,gazuz09,candelier09}, where a strong
pulling force is applied to a single particle, forcing it out of its
cage, thereby possibly melting the glass.

Another system far from equilibrium are athermal packings of particles
\cite{cates+wittmer98,picaciamarra+nicodemi10}, undergoing a jamming
transition at a critical packing fraction. Many of the properties
close to the jamming point resemble those of fluids at the glass
transitions. This observation is at the heart of the jamming diagram,
where the glass transition in thermal systems and the jamming
transition are part of a larger parameter space
\cite{liu+nagel98,ohern+silbert03}.  

Since granular particles are too large to be thermally activated, one
necessarily needs a driving force to keep the grains in motion for
extended periods of time. While in nature, gravity is probably the
most important driving force \cite{iverson97}, experimentalists have
devised quite a few methods of fluidisation. The list includes shaking
\cite{prevost+egolf02}, electrostatic
\cite{aranson+olafsen02,kohlstedt+snezhko05} or magnetic
\cite{kohlstedt+snezhko05,maass+isert08} excitation and fluidization
by air \cite{ojha+lemieux04,abate+durian05} or water
\cite{schroeter+goldman05}.

\begin{figure}[t]
  \centering
  \includegraphics[width=.48\textwidth]{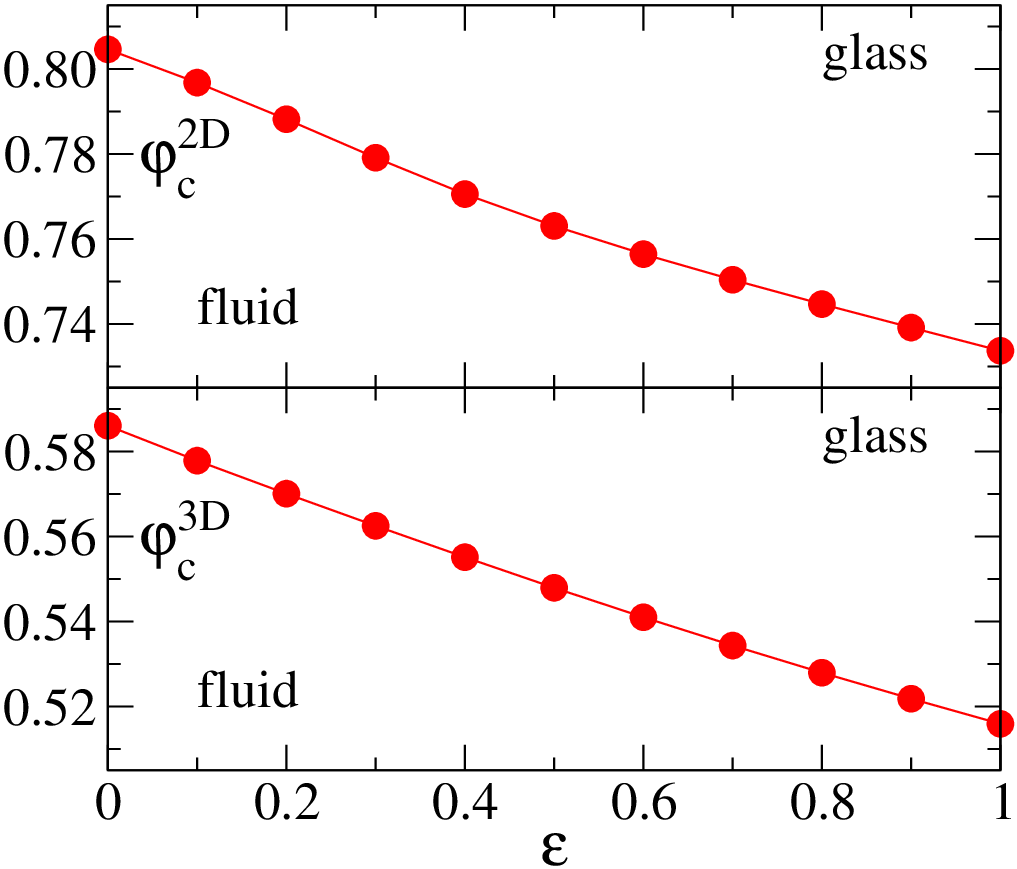}
  \caption{Critical packing fraction, $\varphi_c$, separating the
    fluid from the glassy state of driven granular particles as a
    function of the coefficient of restitution $\varepsilon$ for space
    dimension $D=2$ (top) and $D=3$ (bottom).}
  \label{fig:jphase}
\end{figure}

We have recently investigated the possibility of a glass transition in
driven granular fluids. In two publications
\cite{kranz+sperl10,sperl+kranz12}, henceforth referred to as I and
II, we have demonstrated that mode coupling theory (MCT) can be
generalized to the far from equilibrium stationary state of a granular
fluid. In particular we found a granular glass transition for all
degrees of dissipation, accompanied by the common signatures of a
dense fluid close to the glass transition. Here, a careful derivation
of the granular mode coupling equations is presented and the
consequences of MCT are worked out in detail. We furthermore extend
our previous analysis to two-dimensional systems, which are realized
in many experiments on granular matter
\cite{abate+durian06,reis+ingale07,keys+abate07,prevost+egolf02,abate+durian05}. The
resulting glass transition diagram is shown in Fig.~\ref{fig:jphase}
in the plane spanned by packing fraction, $\varphi$, and coefficient
of restitution, $\varepsilon$.

The dissipative interactions of the granular particles imply two
primary consequences. First, while the dynamics of particles in
thermal equilibrium is microscopically time reversal invariant, the
symmetry under time reversal is broken for granular dynamics. Second,
there is no natural equilibrium reference state like for the sheared
colloids~\cite{fuchs+cates02,fuchs+cates09}, were the fluid can be
thought of as being driven out of equilibrium by the optional external
driving force. In the granular system, the driving force is required
to maintain a stationary state with more than transient dynamics.

The paper is organized as follows. In Sec.~\ref{sec:model} we define
the model of a driven granular fluid and introduce the microscopic
dynamics in Sec.~\ref{sec:micr-descr}. In
Sec.~\ref{sec:gran-glass-trans} we derive the MCT equations for the
coherent scattering function, $\phi(q,t)$, discuss the asymptotic
correlations $f_q := \phi(q,t\to\infty)$, used as an order parameter
to locate the glass transition, and analyze the dynamics close to the
glass transition. The MCT equations for the incoherent scattering
function, $\phi_s(q,t)$, of a tagged particle are derived in
Sec.~\ref{sec:tagg-part-dynam}. In Sec.~\ref{sec:discussion} we
discuss our results in a broader context and conclude with a number of
perspectives for future work in Sec.~\ref{sec:conclusion}.

\section{Model}
\label{sec:model}

\subsection{Inelastic Hard Spheres}
\label{sec:inel-hard-spher}

The granular fluid is modeled as a monodisperse system of $N$ smooth
inelastic hard spheres (in dimension $D=3$) or disks (in $D=2$) of
radius $a$ and mass $m=1$ in a volume $V=L^D$. We consider the
thermodynamic limit $N,V\to\infty$ such that the density $n=N/V$
remains finite. Dissipation is introduced through a constant
coefficient of normal restitution $\varepsilon\in[0,1]$ that augments
the law of reflection \cite{haff83},
\begin{equation}
  \label{eq:1}
  \uvec r_{12}\cdot\vec v'_{12} = -\varepsilon\,\uvec r_{12}\cdot\vec v_{12},
\end{equation}
where $\vec v_{12} = \vec v_1 - \vec v_2$ is the relative velocity and
$\uvec r_{12}$ is the unit vector pointing from the center of particle
2 to particle 1. The prime indicates post-collisional quantities.

\subsection{Stochastic Driving Force}
\label{sec:stoch-driv-force}

The driving force is implemented as an external random force,
\begin{equation}
  \label{eq:2}
  \vec v'_i(t) = \vec v_i(t) + \sqrt{P_D}\,\vec\xi_i(t),
\end{equation}
where $P_D$ is the driving power. The $\xi_i^{\alpha}$,
$\alpha=1,\ldots,D$ are Gaussian random variables with zero mean and
variance,
\begin{equation}
  \label{eq:3}
  \avr{\xi_i^{\alpha}(t)\xi_j^{\beta}(t')}_{\xi} 
  = [\delta_{ij} - \delta_{\pi(i),j}]\delta^{\alpha\beta}\delta(t-t'),
\end{equation}
where $\pi(i) = \argmin_k\{|\vec r_i-\vec r_k|\ge\ell_D\}$ yields the
index of the particle that is closest to particle $i$ but at least a
given distance, $\ell_D$, away. Thereby, the external force does not
destroy momentum conservation on length scales $\ell\gtrsim\ell_D$. We
choose $\ell_D$ on the order of a mean particle separation.

\subsection{The Granular Fluid}
\label{sec:granular-fluid}

For undriven granular fluids, it is known that the homogeneous cooling
state is unstable to shear- and eventually to density fluctuations
\cite{mcnamara93,goldhirsch+zanetti93}. In fact, the particles form
extremely dense clusters. No such clustering instability is predicted
\cite{mcnamara93}, and indeed observed, for the randomly driven
fluid. Consequently, we assume that the fluid is macroscopically
homogeneous and isotropic. This implies that all spatial two-point
correlation functions $C(\vec r, \vec r')$ are functions of the
distance $|\vec r-\vec r'|$ only. In the stationary state, the system
is also time translation invariant, implying that time dependent
correlation functions are only functions of time differences.

Macroscopically, the fluid is fully characterized by the packing
fraction, $\varphi$ (where $\varphi = 4\pi na^3/3$ in $D=3$ and
$\varphi = \pi na^2$ in $D=2$), the coefficient of restitution,
$\varepsilon$, and the driving power, $P_D$. A more conventional
description is given in terms of the packing fraction and the granular
temperature $T = T(\varphi,\varepsilon,P_D) = \frac{1}{DN}\sum_i\vec
v_i^2$. The latter is given by the balance between the driving power,
$P_D$, and the energy loss through the inelastic collisions,
$P_{\varepsilon} := \Pi_{\varepsilon}(\varphi,\varepsilon)\omega_cT$,
where $\Pi_{\varepsilon}(\varphi,\varepsilon)
\simeq\Pi_{\varepsilon,E}(\varepsilon) = (1-\varepsilon^2)/D$ in a
mean field approximation \cite{haff83}.

The collision frequency $\omega_c\propto\sqrt T$ is the only time
scale of the system. Thus, changing the granular temperature only
changes the time scale of the system. To keep the discussion more
transparent, we refrain from using the freedom to set $T=1$ but keep
in mind that the qualitative behavior of the system is independent of
the temperature $T>0$.

\section{Microscopic Description}
\label{sec:micr-descr}

\subsection{Observables}
\label{sec:observables}

The two relevant observables discussed in the following are the
density field, $\rho(\vec r, t)$, and the current density, $\vec
j(\vec r, t)$, with the following microscopic definitions:
\begin{subequations}
\begin{align}
  \label{eq:7}
  \rho(\vec r, t) &= \frac1N\sum_i\delta(\vec r - \vec r_i(t)),\\
  \vec j(\vec r, t) &= \frac1N\sum_i\vec v_i(t)\delta(\vec r - \vec r_i(t)).
\end{align}
\end{subequations}
We will use the spatial Fourier transforms $\rho_{\vec
  k}(t) = \FT[\rho](t)$ and the longitudinal current $j_{\vec k}^L(t) =
\uvec k\cdot\FT[\vec j](t)$. The corresponding tagged particle
quantities are given as
\begin{subequations}
\begin{align}
  \label{eq:8}
  \rho^s(\vec r, t) &= \delta(\vec r - \vec r_s(t)),\\
  \vec j^s(\vec r, t) &= \vec v_s\delta(\vec r - \vec r_s(t)).
\end{align}
\end{subequations}

\subsection{Dynamics}
\label{sec:dynamics}

The (forward in time) pseudo Liouville operator $\Lv_+$ describes the
time evolution of a microscopic observable $A$, i.e., $i\Lv_+A =
dA/dt$, according to the dynamics specified above
\cite{altenberger75}. It is given as the sum of three parts,
\begin{equation}
  \label{eq:4}
  \Lv_+ = \Lv_0 + \sum_{j<k}\mathcal T_{jk}^+ + \mathcal L_D^+,
\end{equation}
which are in turn: (i) The free streaming operator $i\Lv_0 =
\sum_j\vec v_j\cdot\nabla_j$. (ii) The collision operator,
\begin{equation}
  \label{eq:5}
  i\mathcal T_{jk}^+ = -(\uvec r_{jk}\cdot\vec v_{jk})
  \Theta(-\uvec r_{jk}\cdot\vec v_{jk})\delta(r_{jk} - 2a)(b_{jk}^+ - 1),
\end{equation}
where $\Theta(x)$ denotes the Heaviside step function and the operator
$b_{jk}^+$ implements the inelastic collision rule
\cite{aspelmeier+huthmann01} and (iii) the driving operator
\footnote{This is just an unusual rendering of It\=o's Lemma
  \cite{oksendal03}},
\begin{equation}
  \label{eq:6}
  \begin{aligned}
    i\Lv_D^+ = &\sqrt{P_D}\sum_j\vec\xi_j(t)\cdot\left(
      \frac{\partial}{\partial\vec v_j} - \frac{\partial}{\partial\vec v_{\pi(j)}}
    \right)\\
    &+ P_D\sum_j\frac{\partial^2}{\partial\vec v_j^2}.
  \end{aligned}
\end{equation}

With the binary collision expansion \cite{ernst+dorfman69}, formal
power series of the pseudo Liouville operator can be defined. In
particular, this allows to write the propagator $\U(t) =
\exp(it\Lv_+)$ in terms of an exponential operator $\exp(it\Lv_+) :=
\sum_n(it\Lv_+)^n/n!$. The Laplace transformed propagator
$\hat\U(s)\equiv\LT[\U](s) = (s - \Lv_+)^{-1}$ is then also defined as
a power series \footnote{We use the convention $\hat f(s) = \LT[f](s)
  = i\int_0^{\infty}f(t)e^{-ist}dt$.}.

The starting point for the derivation of equations of motion is an
operator identity that is most concisely expressed in the Laplace
domain,
\begin{equation}
  \label{eq:12}
  \P\hat\U(s)\P = [s - \mathsf\Omega - \hat{\mathsf M}(s)]^{-1},
\end{equation}
where $\mathsf{\Omega} = \P\Lv_+\P$,
\begin{equation}
  \label{eq:10}
  \mathsf M(t) = \P\Lv_+\Q\exp(it\Q\Lv_+\Q)\Q\Lv_+\P,
\end{equation}
and $\P=\P^2,\Q=1-\P$ are projection operators \cite{mori65b}.

\subsection{Phase Space Averages}
\label{sec:phase-space-averages}

Our approach starts from the microscopic description of the particle
dynamics in terms of the location in phase space, $\Gamma := (\vec
r_1,\vec v_1,\ldots\vec r_N,\vec v_N)$ and the trajectory of the
external driving force $\Xi_t =
(\vec\xi_1(t),\ldots,\vec\xi_N(t))$. Macroscopic observables $\tilde
A(\vec r, t| \Xi_t) := \avr{a(\vec r, t|\Gamma, \Xi_t)}_{\Gamma}
\equiv \int d\Gamma\varrho(\Gamma,t)a(\vec r, t|\Gamma, \Xi_t)$ are
then introduced as expectation values with respect to the phase space
distribution function $\varrho(\Gamma,t)$ of the microscopic
variables. Here, we restrict ourselves to the stationary state, where
the distribution function is time independent, $\varrho(\Gamma, t)
\equiv \varrho(\Gamma)$.

In contrast to fluids in equilibrium, no analytical expression for the
stationary phase space distribution of driven granular fluids is known
so far. Therefore we have to make a few assumptions to evaluate the
expectation values. First of all we assume that positions and
velocities are uncorrelated, $\varrho(\Gamma) =
\varrho_r(\{\vec r_i\})\varrho_v(\{\vec v_i\})$. Moreover, we assume
that the velocity distribution factorizes into a product of one
particle distribution functions, $\varrho_v(\{\vec v_i\}) =
\prod_i\varrho_1(\vec v_i)$. All we need to know about $\varrho_1(\vec
v)$ is that it has a vanishing first moment, $\int d^Dv\,\vec
v\varrho_1(\vec v) = 0$ and a finite second moment, $\int
d^Dv\,v^2\varrho_1(\vec v) = DT < \infty$. The spatial distribution
function, $\varrho_r(\{\vec r_i\})$, enters the theory via static
correlation function, as will be discusssed below.

A common expectation in disordered systems is, that the macroscopic
expectation values should be self averaging, i.e., independent of any
specific disorder realization \cite{brout59}. In our model, this
applies to the stochastic driving force $\Xi_t$, i.e., we define
macroscopic observables as averages over all realizations of the
driving, $A(\vec r, t) = \avr{a(\vec r, t)} := \avr{\tilde A(\vec r,
  t|\Xi_t)}_{\xi} \equiv\int d\Xi_tP(\Xi_t)\tilde A(\vec r,
t|\Xi_t)$. Here, $P(\Xi_t)$ is the distribution of the random forces.
Averages over pairs of observables define a scalar product, $\avr{A|B}
:= \avr{A^*B}$ where $A^*$ denotes the complex conjugate of $A$.

Given the definition of a scaler product, we can formally introduce
the adjoint Liouville operator, $\Lv_+^{\dagger}$, via the relation
$\avr{\Lv_+^{\dagger}a|b} = \avr{a|\Lv_+b}$. For elastic hard spheres
in thermal equilibrium, it can be shown that detailed balance implies
$\Lv_+^{\dagger}(\varepsilon=1) = \Lv_-(\varepsilon=1)$, where $\Lv_-$
is the Liouville operator describing time reversed dynamics. This
relation does not hold for inelastic collisions which are discussed
here. In the present context, an explicit expression for
$\Lv_+^{\dagger}$ is not needed and hence will be given elsewhere.

As both the hard sphere interactions and the driving force (by
construction) conserve momentum, this will also be reflected in the
matrix elements $\avr{a_1(\vec k_1)\cdots a_n(\vec k_n)|b_1(\vec
  p_1)\cdots b_m(\vec p_m)} \propto \delta_{\sum\vec k_i,\sum\vec
  p_i}$ in the form of a selection rule.

The central quantities in the following will be the coherent
scattering function
\begin{subequations}
  \begin{equation}
    \label{eq:92}
    \phi(q,t) := N\avr{\rho_{\vec q}(\tau)|\rho_{\vec q}(\tau + t)}/S_q,
  \end{equation}
  where $S_q := N\avr{\rho_{\vec q}|\rho_{\vec q}}$ is the static
  structure factor, and the incoherent scattering function
  \begin{equation}
    \label{eq:93}
    \phi_s(q,t) := \avr{\rho_{\vec q}^s(\tau)|\rho_{\vec
        q}^s(\tau+t)}.
  \end{equation}
\end{subequations}
In general, all macroscopic quantities will be functions of the
coefficient of restitution $\varepsilon$. To reduce clutter, we will
suppress this dependence.

\subsection{Static Structure Factors}
\label{sec:structure_factors}

Below, we will treat the static structure functions as a known
input. Hence, we need $S_q = S_q(\varphi, \varepsilon)$ for a range of
densities, $\varphi$ around where we expect the critical density to
be. Lacking good quality data for $S_q(\varphi
\sim\varphi_c(\varepsilon), \varepsilon)$, let alone reliable
theoretical predictions for this quantity, we use preliminary results
that $S_q(\varphi, \varepsilon)$ only weakly depends on the
coefficient of restitution $\varepsilon$ and approximate $S_q(\varphi,
\varepsilon) \approx S_q(\varphi,\varepsilon=1)$ by their elastic
counterparts: In $D=3$ we use the Percus-Yevick (PY) equation
\cite{percus+yevick58} for elastic hard spheres in thermal equilibrium
\cite{ashcroft+lekner66}, except for the pair correlation function at
contact which is better approximated by the Carnahan-Starling
expression \cite{carnahan+starling69}. In $D=2$ we use the Baus-Colot
(BC) equation \cite{baus87} throughout.

\section{The Granular Glass Transition}
\label{sec:gran-glass-trans}

\subsection{Equations of Motion}
\label{sec:equations-motion}

Let us introduce the following microscopic state vector $\vec a_{\vec
  q} = \sqrt N(\rho_{\vec q}/\sqrt{S_q}, j_{\vec q}^L/\sqrt T)$. Then
the coherent scattering function $\phi(q,t)$ is given as one element
of the matrix of correlators $\Phi(q,t) = \avr{\vec a_{\vec
    q}|\mathsf{U}(t)\vec a_{\vec q}}$.

With the help of Eq.~\eqref{eq:12} and the projectors
\begin{equation}
  \label{eq:11}
  \P_c = N\sum_{\vec q}\ket{\rho_{\vec q}}\bra{\rho_{\vec q}}/S_q
  + N\sum_{\vec q}\ket{j_{\vec q}^L}\bra{j_{\vec q}^L}/T,
\end{equation}
$\Q_c = 1 - \P_c$, one finds
\begin{equation}
  \label{eq:14}
  \hat\Phi^{-1}(q,s) =
  \begin{pmatrix}
    s & -\Omega_{\rho j}(q)[1 + \hat L(q,s)]\\
    -\Omega_{j\rho}(q) & s - \nu_q - \hat M(q,s)
  \end{pmatrix},
\end{equation}
while $\Omega_{\rho\rho} \propto \avr{\rho_{\vec q}|\Lv_+\rho_{\vec
    q}} = q\avr{\rho_{\vec q}|j_{\vec q}^L} = 0$ due to parity. The
other entries of the frequency matrix $\Omega_{ab} = \avr{\vec a_{\vec
    q}|\mathsf{\Omega}\vec a_{\vec q}}$ are nonzero [Note that $\nu_q
= \Omega_{jj}(q)$].

The memory kernels are formally given as
\begin{subequations}
\begin{align}
  \label{eq:9}
  M(q,t) &= N\avr{F_{\vec q}^{\dagger}\tilde{\mathsf U}(t)F_{\vec q}}/T,\\
  L(q,t) &= \avr{J_{\vec q}^{\dagger}\tilde{\mathsf U}(t)F_{\vec q}}/q
  \avr{j_{\vec q}^L|j_{\vec q}^L},
\end{align}
\end{subequations}
where $\tilde{\mathsf U}(t) = \exp(it\Q_c\Lv_+\Q_c)$ is a modified
propagator. There are two fluctuating forces, $F_{\vec q} =
\Q_c\Lv_+j_{\vec q}^L$, and $F_{\vec q}^{\dagger} =
\Q_c\Lv_+^{\dagger}j_{\vec q}^L$, and at this point we can not rule out
that there is a nonzero fluctuating current, $J_{\vec q}^{\dagger} =
\Q_c\Lv_+^{\dagger}\rho_{\vec q}$, while $J_{\vec q} =
\Q_c\Lv_+\rho_{\vec q} = q\Q_c j_{\vec q}^L = 0$. In the elastic limit
$F_{\vec q}^{\dagger} = F_{\vec q}$ and $J_{\vec q}^{\dagger} =
J_{\vec q} = 0$ holds and therefore $L(q,t)$ vanishes.

In the Laplace domain, the coherent scattering function is thus given
as
\begin{equation}
  \label{eq:13}
  \hat\phi^{-1}(q,s) 
  = s - \frac{\Omega_q^2[1 + \hat L(q,s)]}{s - \nu_q -
    \hat M(q,s)},
\end{equation}
where $\Omega_q^2 = \Omega_{\rho j}\Omega_{j\rho}$, or, equivalently, in
the time domain as the solution of the equation of motion,
\begin{equation}
  \label{eq:18}
  \begin{aligned}
    \ddot\phi(q,t) &+ \nu_q\dot\phi(q,t) + \Omega_q^2\phi(q,t)\\
    &+ \Omega_q^2\int_0^t\!d\tau\,m(q,t-\tau)\dot\phi(q,\tau)\\
    &+ \Omega_q^2\int_0^t\!d\tau\,L(q,t-\tau)\phi(q,\tau) = 0,
  \end{aligned}
\end{equation}
where $m(q,t) = M(q,t)/\Omega_q^2$ and the initial conditions are
$\dot\phi(q,t=0) = 0$ and $\phi(q,t = 0) = 1$. To proceed, we need to
find approximate expressions for the memory kernels. Before we come to
the mode coupling approximation, we discuss the simpler assumption
that $M(q,t) = L(q,t) = 0$.

\subsection{Sound Waves}
\label{sec:sound-waves}

The linear equation of motion
\begin{equation}
  \label{eq:19}
  \ddot\phi(q,t) + \nu_q\dot\phi(q,t) + \Omega_q^2\phi(q,t) = 0
\end{equation}
describes damped sound waves $\phi(q,t) = e^{-\nu_qt/2}\cos(C_qqt)$.

Sound damping due to collisions, as described by the second term in
Eq.~\eqref{eq:19},
\begin{equation}
  \label{eq:200}
  \nu_q = N\avr{j_{\vec q}^L|\Lv_+j_{\vec q}^L}/T,
\end{equation}
can be evaluated in the Enskog approximation. The calculation for two
dimensions is shown in appendix~\ref{sec:freq-omeg-two}, yielding
\begin{subequations}
\begin{equation}
  \label{eq:32a}
  \nu_q = \frac{1+\varepsilon}{2}\omega_E[1 + 2J_0''(2aq)]
  \quad\quad\text{in $D=2$,}
\end{equation}
where $J_0(x)$ is the zeroth order Bessel function \cite{gradshteyn00} and
the double prime denotes the second derivative with respect to the
argument. The result in three dimensions is
known~\cite{leutheusser82a},
\begin{equation}
  \label{eq:32b}
  \nu_q = \frac{1+\varepsilon}{3}\omega_E[1 + 3j_0''(2aq)]
  \quad\quad\text{in $D=3$,}
\end{equation}
\end{subequations}
where $j_0(x) = \sin(x)/x$ is the zeroth order spherical Bessel
function \footnote{A factor of $2/3$ was missing in I \& II. This has
  no influence on the results discussed there.}. The Enskog collision
frequency $\omega_E = 2^DD(\varphi\chi/2a)\sqrt{T/\pi}$ is given in
terms of the contact value, $\chi$, of the pair correlation function
\cite{hansen+mcdonald06}.

The speed of sound is given in the long wavelength limit by $ C_q^2 =
\Omega_q^2/q^2= \Omega_{\rho j}\Omega_{j\rho}/q^2$.  One finds,
\begin{subequations}
\begin{equation}
  \label{eq:94}
  \begin{aligned}
    \Omega_{j\rho} &= N\avr{j_{\vec q}^L|\Lv_+\rho_{\vec q}}/\sqrt{TS_q}\\ 
    &= qS_{\ell\ell}(q)\sqrt{T/S_q}
  \end{aligned}
\end{equation}
with the longitudinal current correlator $S_{\ell\ell}(q) :=
N\avr{j_{\vec q}^L|j_{\vec q}^L}/T$. The calculation
(cf. appendix~\ref{sec:freq-omeg-j}) of
\begin{equation}
  \label{eq:15}
  \begin{aligned}
    \Omega_{\rho j} &= \frac{N}{\sqrt{TS_q}}\avr{\rho_{\vec
        q}|\Lv_+j_{\vec q}^L}\\
    &\approx q\sqrt{T/S_q}\left( \frac{1 + \varepsilon}{2} +
      \frac{1-\varepsilon}{2}S_q \right)
  \end{aligned}
\end{equation}
\end{subequations}
uses the approximate granular Yvon-Born-Green (YBG) relation
(cf. appendix~\ref{sec:granular-born-green}). Combining these results,
we find that the long wavelength speed of sound,
\begin{equation}
  \label{eq:20}
  C_q^2 = T\frac{S_{\ell\ell}(q)}{S_q}\left(
    \frac{1 + \varepsilon}{2} + \frac{1-\varepsilon}{2}S_q
  \right),
\end{equation}
is reduced for the dissipative driven fluid compared to a fluid of
elastic hard spheres in thermal equilibrium. The sound damping,
$\nu_q/2q^2$, on the other hand decreases with increasing dissipation.

Alternatively, the speed of sound can be given as $C_q^2 =
S_{\ell\ell}(q)/n\kappa_q^{\mathrm{eff}}$, where the effective
compressibility $\kappa_q^{\mathrm{eff}}$ is defined in a form,
\begin{equation}
  \label{eq:77}
  \frac{1}{\kappa_q^{\mathrm{eff}}} = \frac{1}{\kappa_q^0} 
  + \frac{1-\varepsilon}{2}Tn^2c_q,
\end{equation}
reminiscent of a random phase approximation \cite{schweizer+curro88}.
Here, $\kappa_q^0 = S_q/nT$ is the compressibility of a fictitious
elastic hard sphere system with a structure factor $S_q$ and $c_q$ is
the direct correlation function \cite{hansen+mcdonald06}.

\begin{figure}[t]
  \centering
  \includegraphics[width=.48\textwidth]{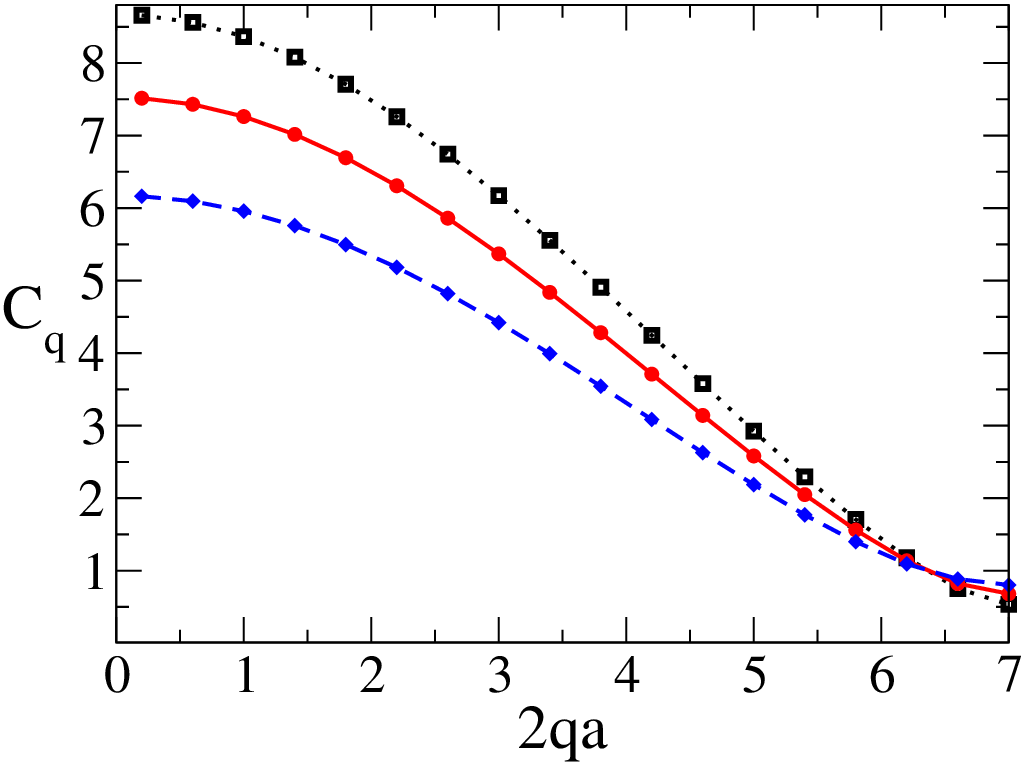}
  \caption{Speed of sound, $C_q$, according to Eq.~(\ref{eq:20}) 
    as a function of wave
    number, $q$, in 3D for packing fraction $\varphi=0.516$ and
    coefficient of restitution $\varepsilon = 1.0$ (squares), 0.5
    (filled circles), and 0.0 (diamonds).}
  \label{fig:Cqeps}
\end{figure}

We close this section with a few remarks. First, in a thermal fluid
the expression for the sound velocity simplifies, because
$S_{\ell\ell}(q)\equiv1$ due to molecular chaos \cite{boon+yip92}. In
a granular fluid, $S_{\ell\ell}(q)$ is actually found to be wave
number dependent. Preliminary results indicate $S_{\ell\ell}(q\to0) <
1$. Second, from $\Omega_{\rho j}\ne\Omega_{j\rho}$ for $\varepsilon <
1$, it can be seen explicitly that the Liouville operator is not self
adjoint. In terms of physical processes, this reflects that the
transition rate for the conversion of density fluctuations into
current fluctuations is not equal to the rate of the reverse
process. Detailed balance, or more general, time reversal invariance
it broken already for the linear equation of motion
(cf. Sec.~\ref{sec:equation-motion} below). Finally, it is known
\cite{hansen+mcdonald06} that Navier-Stokes-order hydrodynamics does
not exist in $D=2$, presumably implying logarithmic corrections to the
sound damping in $D=2$.

In the following, we set $S_{\ell\ell}(q) \equiv 1$. In
Fig.~\ref{fig:Cqeps}, the resulting sound dispersion relations are
shown for packing fraction $\varphi=0.516$ and coefficient of
restitution varying from $\varepsilon = 1$ to 0. The speed of sound
decreases with increasing dissipation in agreement with hydrodynamic
predictions \cite{vannoije+ernst99}.

\subsection{The Mode Coupling Approximation}
\label{sec:mode-coupl-appr}

In the spirit of the equilibrium mode coupling theories
\cite{Fixman62,Kawasaki66,Kadanoff+Swift68,bosse+goetze78b}, we
introduce a second projection operator
\begin{equation}
  \label{eq:22}
  \P_2 = N^2\sum_{\vec k, \vec p}\ket{\rho_{\vec k}\rho_{\vec
      p}}\bra{\rho_{\vec k}\rho_{\vec p}}/S_kS_p
\end{equation}
and approximate the modified propagator as
\begin{equation}
  \label{eq:21}
  \begin{aligned}
    \tilde{\mathsf U}(t) &\approx \P_2\tilde{\mathsf U}(t)\P_2\\
    &\approx N^2\sum_{\vec k, \vec p}\ket{\rho_{\vec k}\rho_{\vec p}}
    \phi(k,t)\phi(p,t)\bra{\rho_{\vec k}\rho_{\vec p}}/S_kS_p,
  \end{aligned}
\end{equation}
where in the second step, a factorization approximation,
$N^2\avr{\rho_{\vec k}\rho_{\vec p}|\tilde{\mathsf{U}}(t)\rho_{\vec
    k}\rho_{\vec p}}/S_kS_p \approx \phi(k,t)\phi(p,t)$, was
employed. Eq.~\eqref{eq:21} is known as the mode coupling
approximation (MCA).

\subsubsection{The MCA of $L$}
\label{sec:mca-l}

We find
\begin{equation}
  \label{eq:23}
  L(q,t) \approx \frac{N}{qT}\sum_{\vec k, \vec p}
  \mathcal U_{\vec q\vec k\vec p}\mathcal W_{\vec q\vec k\vec p}\phi(k,t)\phi(p,t),
\end{equation}
where
\begin{equation}
  \label{eq:24}
  \mathcal U_{\vec q\vec k\vec p} = N\avr{\rho_q|\Lv_+\Q_c\rho_k\rho_p}/S_k
  = 0
\end{equation}
due to parity and $\mathcal W_{\vec q\vec k\vec p}$ is defined
below. Therefore $L(q,t) \equiv 0$ within the mode coupling
approximation.

\subsubsection{The MCA of $M$}
\label{sec:mca-m}

The mode coupling approximation for $M(q,t)$ yields
\begin{equation}
  \label{eq:25}
  M(q,t) \approx \frac NT\sum_{\vec k,\vec p}\mathcal V_{\vec q\vec
    k\vec p}\mathcal W_{\vec q\vec k\vec p}\phi(k,t)\phi(p,t),
\end{equation}
where
\begin{subequations}
\begin{align}
  \label{eq:26}
  \mathcal V_{\vec q\vec k\vec p} 
  &= N\avr{j_{\vec q}^L|\Lv_+\Q_c\rho_{\vec k}\rho_{\vec p}}/S_k,\\
  \mathcal W_{\vec q\vec k\vec p} 
  &= N\avr{\rho_{\vec k}\rho_{\vec p}|\Q_c\Lv_+j_q^L}/S_p.
\end{align}
\end{subequations}

The left vertex is known from the literature \cite{barrat+goetze89}
\begin{equation}
  \label{eq:27}
  \mathcal V_{\vec q\vec k\vec p} 
  = \frac{T}{NS_k}[(\uvec q\cdot\vec k)S_p + (\uvec q\cdot\vec p)S_k -
  qS^{(3)}(\vec k, \vec p)/S_q]\delta_{\vec q,\vec k+\vec p}.\notag
\end{equation}
Customarily, the convolution approximation \cite{jackson+feenberg62},
$S^{(3)}(\vec k,\vec p) \approx S_kS_pS_q$, is applied to yield
\begin{subequations}
\begin{equation}
  \label{eq:28}
  \mathcal V_{\vec q\vec k\vec p} 
  = \frac TNS_p[(\uvec q\cdot\vec k)nc_k + (\uvec q\cdot\vec p)nc_p]
  \delta_{\vec q,\vec k+\vec p}.
\end{equation}

By a nontrivial calculation (see
appendix~\ref{sec:vertex-mathal-w_vec}), we were able to show that the
right vertex is approximately given as
\begin{equation}
  \label{eq:17}
  \mathcal W_{\vec q\vec k\vec p} 
  \approx \frac{1+\varepsilon}{2}\frac TNS_k
  [(\uvec q\cdot\vec k)nc_k + (\uvec q\cdot\vec p)nc_p]
  \delta_{\vec q,\vec k+\vec p},
\end{equation}
\end{subequations}
different from$\mathcal V_{\vec q\vec k\vec p}$.

The physical interpretation of these results for the vertices is, that
(i) the rate of annihilation of pairs of density fluctuations
$\rho_{\vec k},\rho_{\vec p}$ is determined by the static structure of
the fluid, both in an equilibrium fluid and in the driven granular
fluid; (ii) The rate of creation of such density fluctuations is
suppressed by a factor $(1 + \varepsilon)/2$, though, compared to the
rate of creation or to the equivalent rate in an equilibrium fluid. 

The reduced memory kernel in the mode coupling approximation (first
reported in I) then reads
\begin{widetext}
  \begin{equation}
    \label{eq:29}
    m[\phi](q,t) = A_q(\varepsilon)\frac{nS_q}{q^2}\int d^Dk\,S_kS_{|\vec q-\vec k|}
    \left\{[\uvec q\cdot\vec k]c_k + [\uvec q\cdot(\vec q-\vec k)]c_p\right\}^2
    \phi(k,t)\phi(|\vec q-\vec k|,t),
  \end{equation}
\end{widetext}
where
\begin{equation}
  \label{eq:79}
  A^{-1}_q(\varepsilon) = 1 + \frac{1-\varepsilon}{1+\varepsilon}S_q.
\end{equation}
 
\begin{figure}[t]
  \centering
  \includegraphics[width=.48\textwidth]{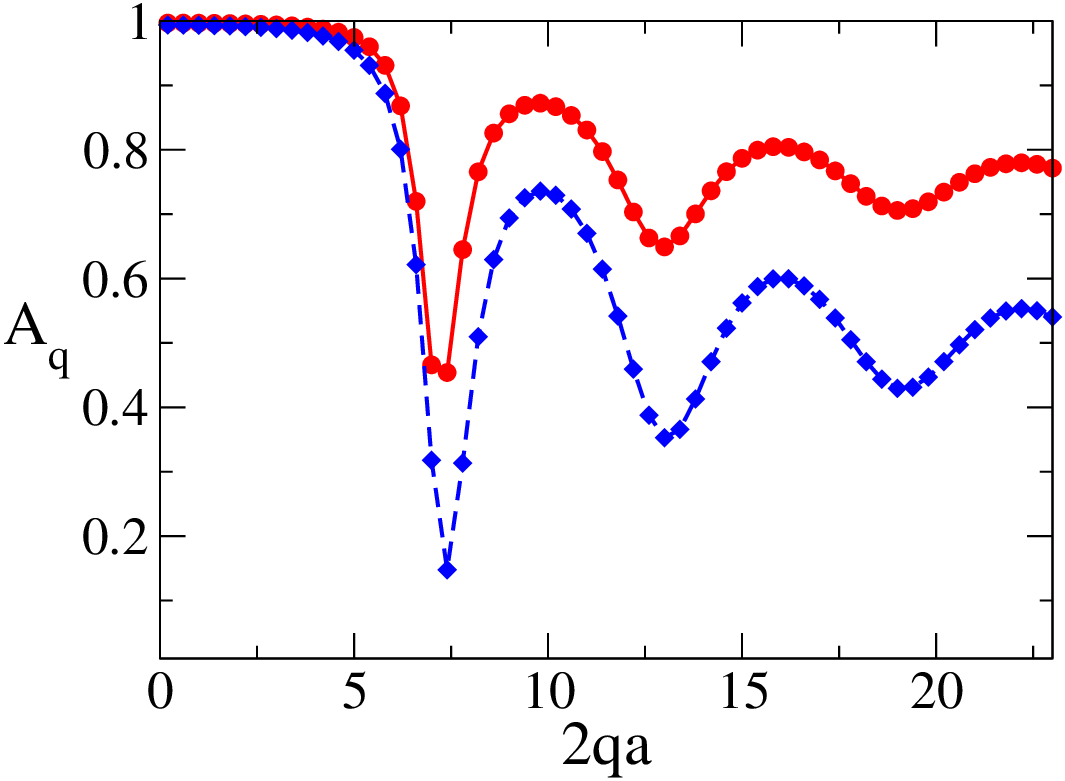}
  \caption{\label{fig:Aqeps} Prefactor of the memory kernel,
    $A_q(\varepsilon)$, as a function of wave number, $q$, for two
    values of the coefficient of restitution, $\varepsilon= 0.5$
    (filled circles) and 0.0 (diamonds), in 3D.  }
\end{figure}

Fig.~\ref{fig:Aqeps} demonstrates the prefactor $A_q(\varepsilon)$
that distinguishes the granular memory functions from the well-known
elastic results where $A_q(\varepsilon=1.0) =1$: For $\varepsilon <
1$, the prefactor exhibits deviations from unity with oscillations
given by the static structure factor. As $A_q(\varepsilon)$ is minimal
for the first peak of the structure factor, i.e. the length scale of
the cage, one concludes that increasing dissipation (decreasing
coefficient of restitution) weakens the cage effect. Compared to the
elastic case, the force acting by the cage onto the particles inside
the cage is smaller as the particles' reflections from each other are
reduced by the influence of dissipation.  While additional changes are
expected by the $\varepsilon$-dependence of the structure factors, the
major difference is encoded in the prefactor $A_q(\varepsilon)$.  The
fact that $A_q(\varepsilon)>0$ for all values of the coefficient of
restitution $\varepsilon$ ensures that the memory kernel remains
positive.

\subsection{The approximate equation of motion and the phase diagram}
\label{sec:appr-equat-moti}

With the mode coupling approximations in place, the equation of motion
\begin{equation}
  \label{eq:30}
  \begin{aligned}
    \ddot\phi(q,t) &+ \nu_q\dot\phi(q,t) + \Omega_q^2\phi(q,t)\\
    &+ \Omega_q^2\int_0^t\!d\tau\,m[\phi](q,t-\tau)\dot\phi(q,\tau) = 0
  \end{aligned}
\end{equation}
turns into a closed equation for the coherent scattering function once
the static structure factor, $S_q$, is known.  This equation of motion
has the same formal structure as the one for the elastic hard sphere
fluid in thermal equilibrium. The viscous term, Eq.~\eqref{eq:200},
decreases with decreasing coefficient of restitution $\varepsilon$;
the speed of sound, Eq.~\eqref{eq:20}, acquires a nontrivial
dependence on the coefficient of restitution as does the memory kernel
$m[\phi]$.

Structural arrest of the grains in a glassy state gives rise to time
persistent density correlations. Hence, we introduce the order
parameter for the glass transition, $f_q :=
\lim_{t\to\infty}\phi(q,t)$. It can
readily be shown that the above equation of motion yields the
following equation for the asymptotic function, $f_q$,
\begin{equation}
  \label{eq:80}
  \frac{f_q}{1-f_q} = m[f](q).
\end{equation}
With the memory kernel being independent of temperature, the order
parameter $f_q$ is also independent of temperature as expected. It can
easily be checked, that $f_q\equiv0$ is always a solution of the above
equation. Studying this equation from a dynamical systems point of
view, one finds that at a critical density $\varphi_c$, the vanishing
solution becomes unstable and a new, stable solution $f_q>0$ appears
discontinuously, signaling the glass transition \cite{goetze09}. The
order parameter at the critical density $\varphi_c$ will be denoted as
$f_q^c$.

Using the structure factors as discussed in
Sec.~\ref{sec:structure_factors}, we find the phase diagrams in
Fig.~\ref{fig:jphase}. For $D=3$ this result was first reported in I,
whereas for $D=2$ this is a new result (for technical parameters,
cf. appendix~\ref{sec:details-numerics}). The order parameter jumps
discontinuously at the critical density $\varphi_c$ as expected
\cite{goetze09} from the type of singularity in Eq.~\eqref{eq:80}.
The evolution of the transition with $\varepsilon$ is remarkably
similar for 2D and 3D, with transition densities increasing from the
elastic case to $\varepsilon = 0.0$ by around 10\% in a roughly linear
fashion.

\subsection{Coherent Dynamics Close to the Glass Transition}
\label{sec:coher-dynam-close}

The full dynamics of density fluctuations is obtained by solving the
MCT equations by iteration (for details, see
Appendix~\ref{sec:details-numerics}).  In Figs.~\ref{fig:coh05_42} and
\ref{fig:coh05_70} we show the coherent scattering function in $D=3$
for several densities, below and above the critical point for
$\varepsilon=0.5$ and two wave numbers $2qa=4.2$ and $2qa=7$. As the
critical point is approached from the fluid side, one observes the
development of a plateau in the coherent scattering
function. Increasing the density above the critical value $\varphi_c$
leads to an increase in the order parameter, $f_q$
\cite{franosch+fuchs97}.

\begin{figure}[t]
  \centering
  \includegraphics[width=.48\textwidth]{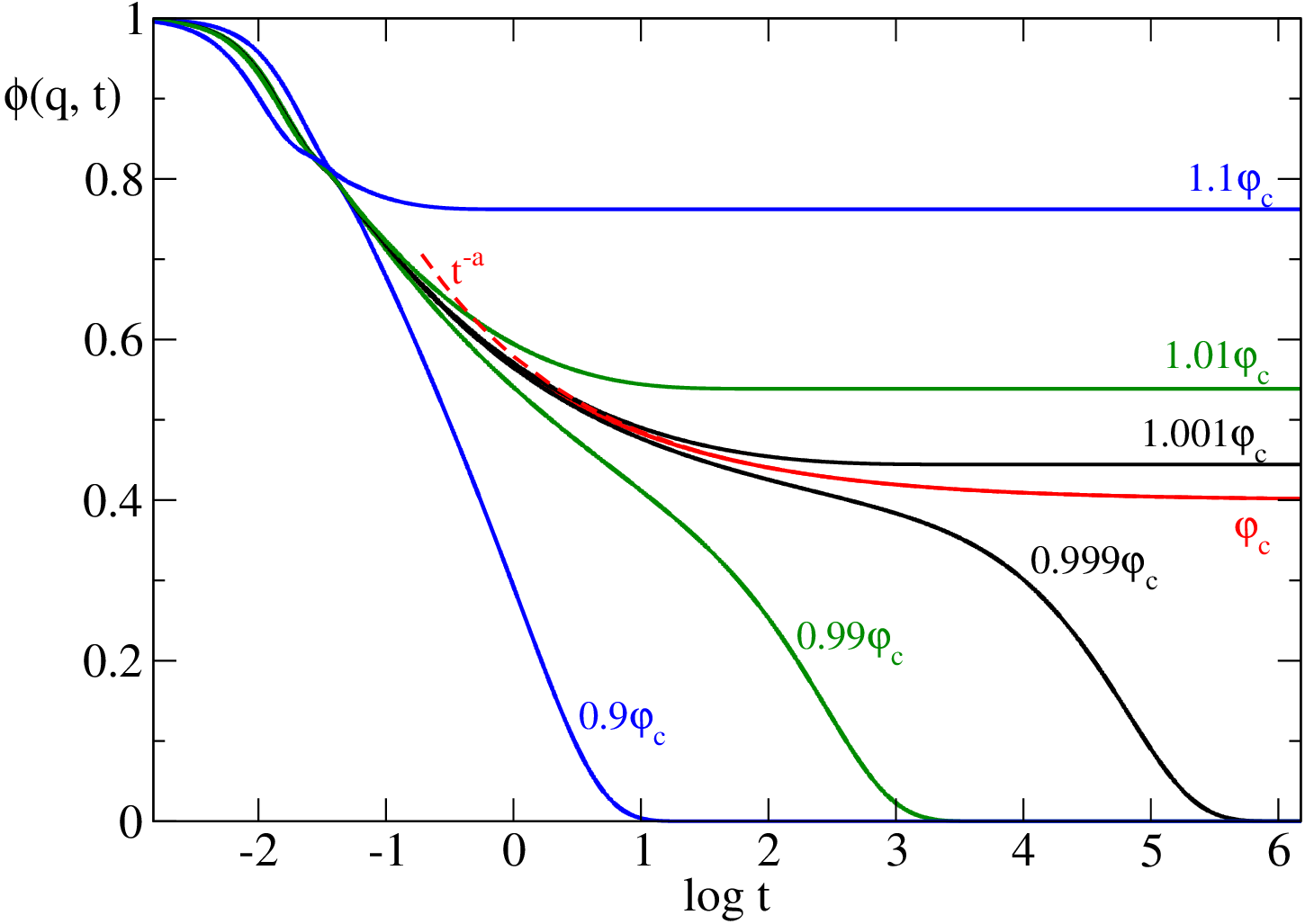}
  \caption{\label{fig:coh05_42} Coherent scattering functions,
    $\phi(q,t)$, as a function of time, $t$, in 3D for wave number $q
    = 4.2/2a$. At the transition point at packing fraction
    $\varphi^c(\varepsilon=0.5) = 0.548$ with a critical glass-form
    factor of $f_q^c=0.400$, and at higher ($1.1\varphi^c$,
    $1.01\varphi^c$, $1.001\varphi^c$) and lower ($0.9\varphi^c$,
    $0.99\varphi^c$, $0.999\varphi^c$) packing fractions. The exponent
    parameter is $\lambda = 0.710$ which yields the critical exponent
    $a = 0.323$, and von Schweidler exponent $b = 0.624$. The critical
    amplitude for $2qa = 4.2$ is $h_q = 0.583$. The critical law
    labeled $t^{-a}$ is shown dashed for a time scale $t_0 = 0.0260$.}
\end{figure}

\begin{figure}[t]
  \centering
  \includegraphics[width=.48\textwidth]{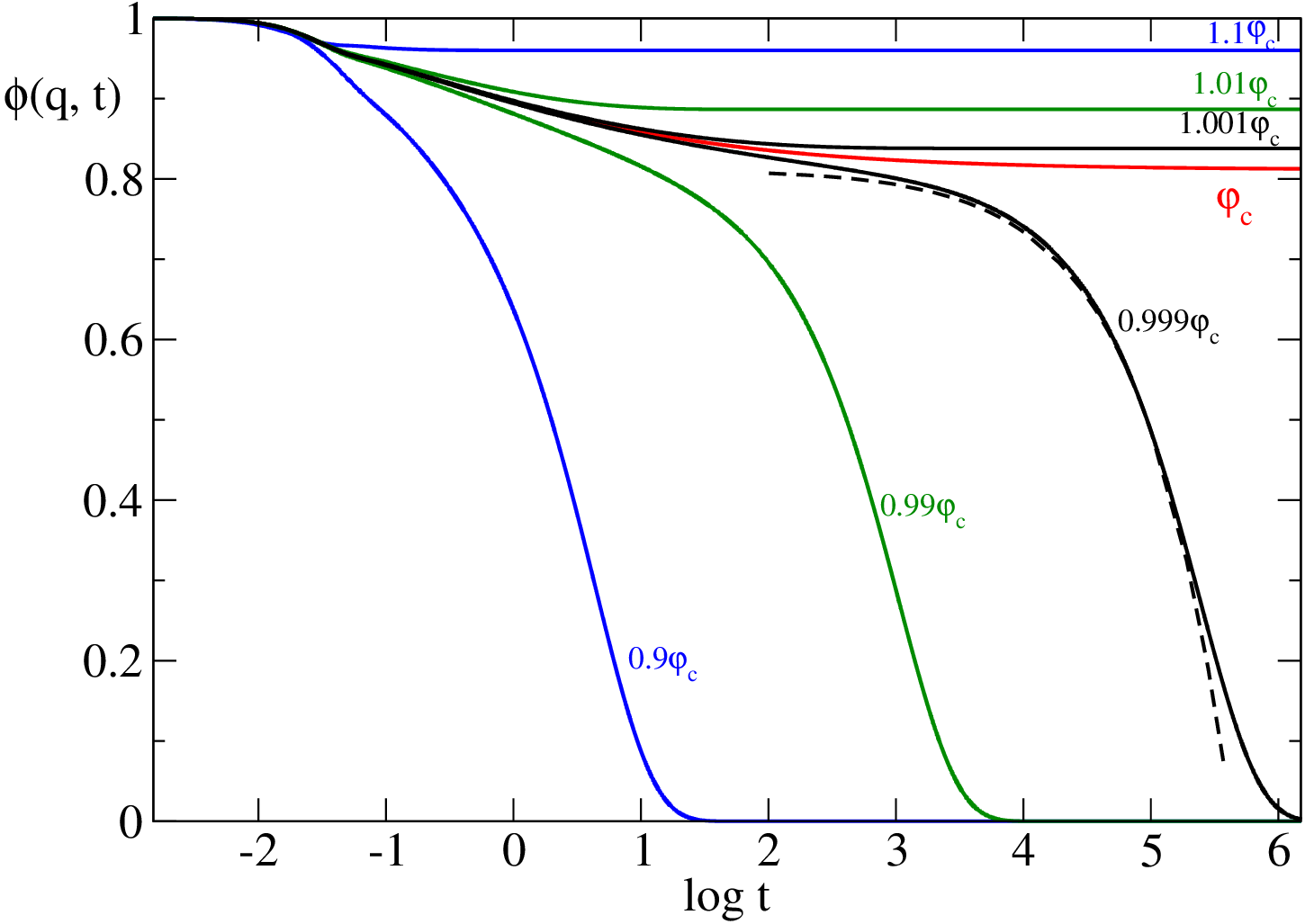}
  \caption{\label{fig:coh05_70} Coherent scattering functions,
    $\phi(q,t)$, as a function of time, $t$, in 3D
    (cf. Fig.~\ref{fig:coh05_42}) for a different wave number $q =
    7/2a$.  The critical amplitude for $2qa = 7$ is $h_q = 0.367$. The
    von Schweidler law is shown dashed for a fitted time scale $\tau =
    1.21\times10^5$.}
\end{figure}

The MCT equations of motion are known to admit scaling solutions at
densities close to the critical density $\varphi_c(\varepsilon=1)$
\cite{goetze09}. As the granular mode coupling equations are formally
identical to those for an equilibrium hard sphere fluid, the scaling
analysis is also closely related. To keep the presentation self
contained, we will summarize the main results of this analysis. The
analogous analysis for the incoherent scattering function discussed
below in Sec.~\ref{sec:tagg-part-dynam} was presented in II.

At small distances $\sigma(\varepsilon) :=
[\varphi_c(\varepsilon)-\varphi]/\varphi_c(\varepsilon)$ to the
critical point, one finds
\begin{equation}
  \label{eq:81}
  \phi(q,t; \sigma) = f_q^c + h_qG_{\sigma}(t),
\end{equation}
where $h_q = h_q(\varepsilon)$ is the critical amplitude and
$G_{\sigma}(t)$ is a scaling function, independent of the wave number.
The scaling function $G_{\sigma}(t)$ can be characterized by a
hierarchy of time scales. The shortest time scale is naturally
provided by the mean time between collisions, $t_0
\equiv\omega_c^{-1}$. The time scale $t_{\sigma}$ for the transition
through the plateau, where $\phi(q, t_{\sigma}; \sigma) = f_q^c$,
diverges at the glass transition as $t_{\sigma} \propto
\sigma^{-\delta}$ where $\delta = 1/2a$. For short but macroscopic
times, in the $\beta$-regime, one finds
\begin{subequations}
  \label{eq:exponent}
  \begin{equation}
    \label{eq:82}
    G_{\sigma}(t) \propto (t/t_0)^{-a},\quad t_0\ll t\le t_{\sigma}.
  \end{equation}
  For larger times, $t \ge t_{\sigma}$ but still well below the so
  called $\alpha$-relaxation time scale $\tau$
  \begin{equation}
    \label{eq:83}
    G_{\sigma}(t) \propto -(t/t_{\sigma})^b,\quad t_{\sigma}\le t \ll \tau.
  \end{equation}
\end{subequations}

Finally, for the largest times the time-density superposition
principle holds, i.e., the coherent scattering functions can be
collapsed on a master curve
\begin{equation}
  \label{eq:84}
  \phi(q,t; \sigma) = \tilde\phi(q, t/\tau(q;\sigma)).
\end{equation}
The time scale $\tau$ diverges at the glass transition, $\tau =
\tau(q; \sigma) \propto \sigma^{-\gamma}$ where $\gamma = (1/2a) +
(1/2b)$.  An empirical choice for the scaling function
$\tilde\phi(q,t)$ is provided by the Kohlrausch-Williams-Watts law
$\tilde\phi_{\mathrm{KWW}}(q, t) \propto \exp(-t^{\beta})$
\cite{williams+watts70}, crossing over to an exponential decay for the
longest times \cite{fuchs94}.

The critical exponents $a,b$ in Eqs.~\eqref{eq:exponent} 
are related to a single exponent parameter
$\lambda=\lambda(D,\varepsilon)$ by the universal relation
\begin{equation}
  \label{eq:85}
  \lambda = \frac{\Gamma^2(1-a)}{\Gamma(1-2a)} 
  = \frac{\Gamma^2(1 + b)}{\Gamma(1 + 2b)},
\end{equation}
where $\Gamma(x)$ is the Euler-Gamma-function. 

\begin{figure}[t]
  \centering
  \includegraphics[width=.45\textwidth]{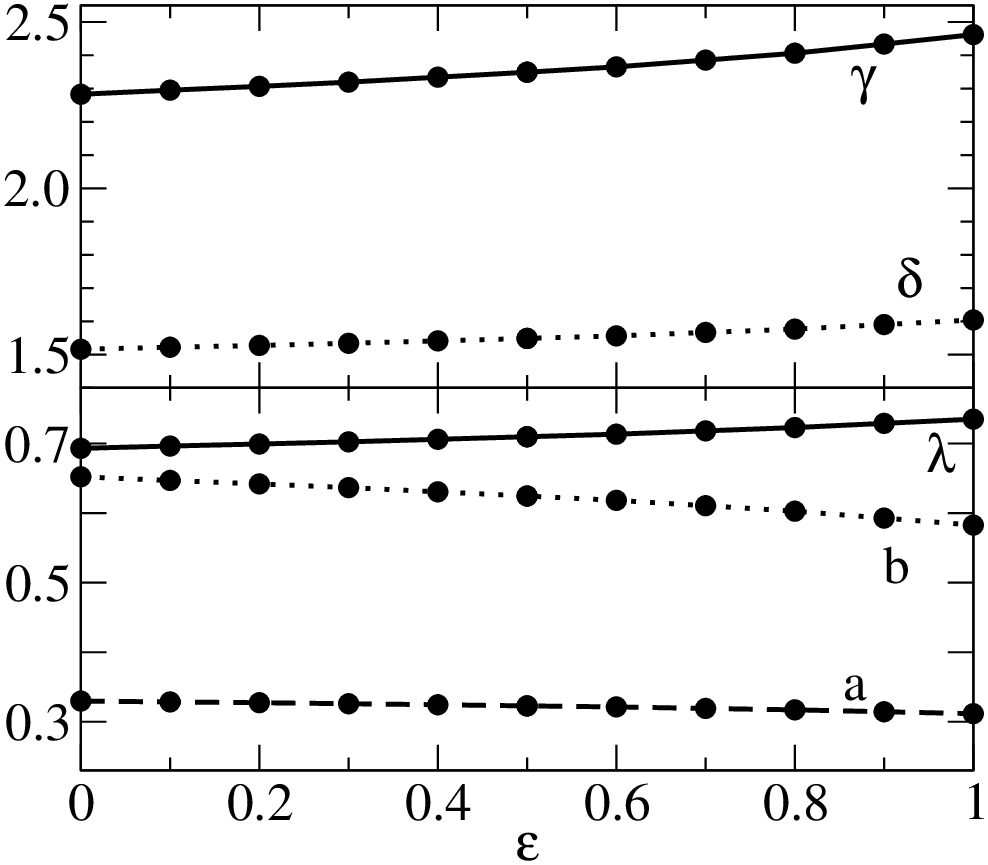}
  \caption{\label{fig:exp3D} Complete set of exponents in 3D as 
    function of the coefficient of restitution $\varepsilon$. The top 
    panel shows the exponents $\delta$ and $\gamma$ for the divergence of 
    the time scales. The lower panel shows the exponents $a$ and $b$
    for the master functions and the exponent parameter $\lambda$.  }
\end{figure}

\begin{figure}[t]
  \centering
  \includegraphics[width=.45\textwidth]{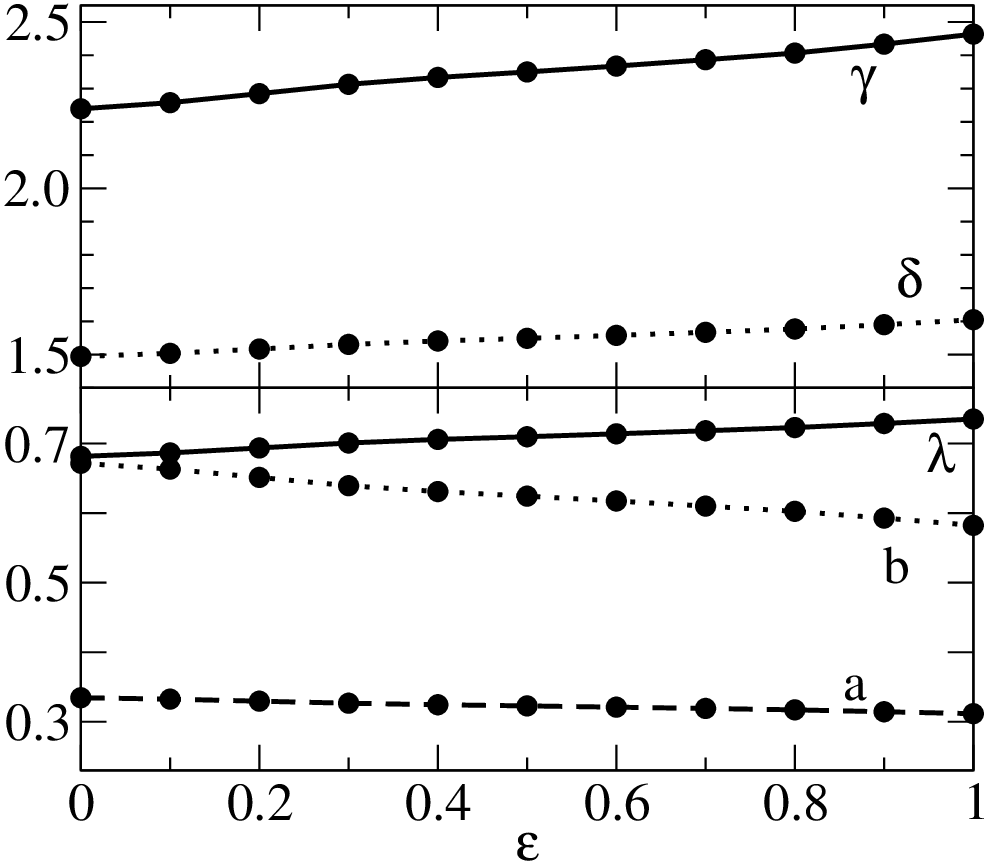}
  \caption{\label{fig:exp2D} Complete set of exponents in 2D as 
    function of the coefficient of restitution $\varepsilon$. The top 
    panel shows the exponents $\delta$ and $\gamma$ for the divergence 
    of the time scales. The lower panel shows the exponents $a$ and $b$
    for the master functions and the exponent parameter $\lambda$.  }
\end{figure}

The sets of exponents shown in Figs.~\ref{fig:exp3D} and
\ref{fig:exp2D} in 3D and 2D, respectively, are functions of the
dimension but differ only slightly between $D=2$ and $D=3$. Overall,
there is a tendency for the exponents in Figs.~\ref{fig:exp3D} and
\ref{fig:exp2D} to show slightly lesser stretching for higher
dissipation, i.e. smaller $\varepsilon$. This may be interpreted as
that the more dissipative and also more strongly driven fluid
experiences less distinctive features in its glassy dynamics. The
divergence in time scales is also expected to be a bit less
pronounced.  While the predicted changes in exponents are most likely
hard to detect in experiments and simulation as absolute numbers, one
should be able to detect the changes in comparison of different
degrees of dissipation. Especially, the master functions should be
comparably sensitive to changes in the exponents $a$ and $b$.

\begin{figure}[t]
  \centering
  \includegraphics[width=.48\textwidth]{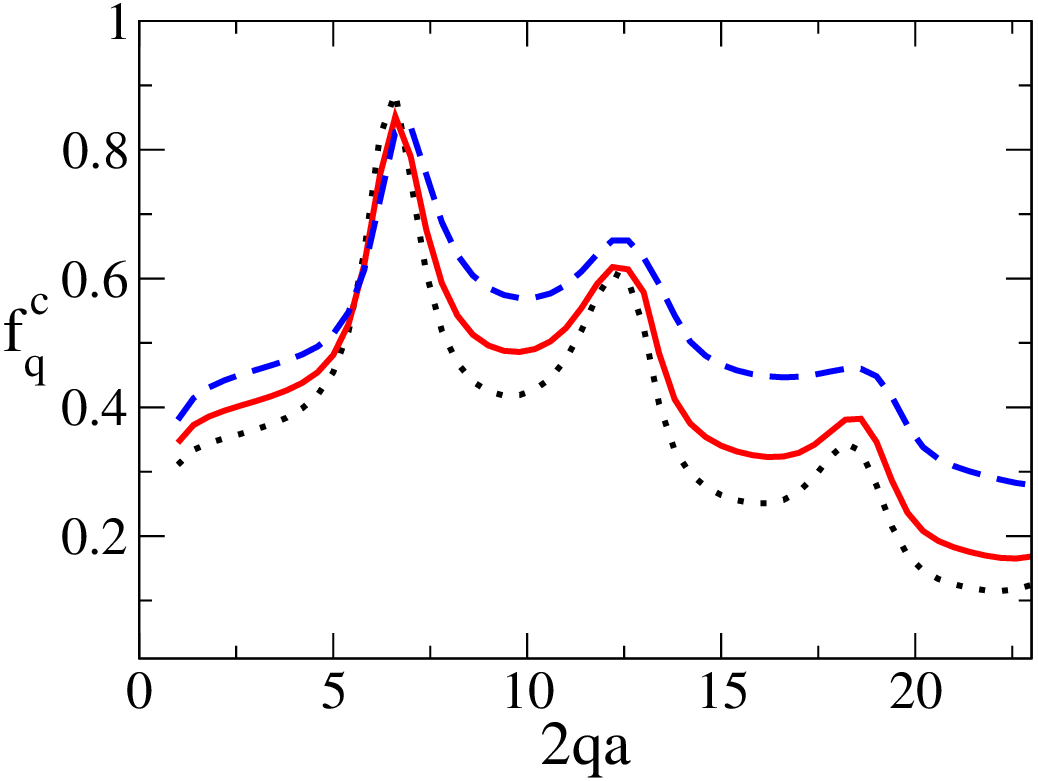}
  \caption{\label{fig:fq2Deps} Critical glass-form factors $f_q^c$ at the 
    transition for $\varepsilon = 1.0$ (dotted curve),
    0.5 (full curve) and 0.0 (dashed curve) in 2D.  }
\end{figure}

\begin{figure}[t]
  \centering
  \includegraphics[width=.48\textwidth]{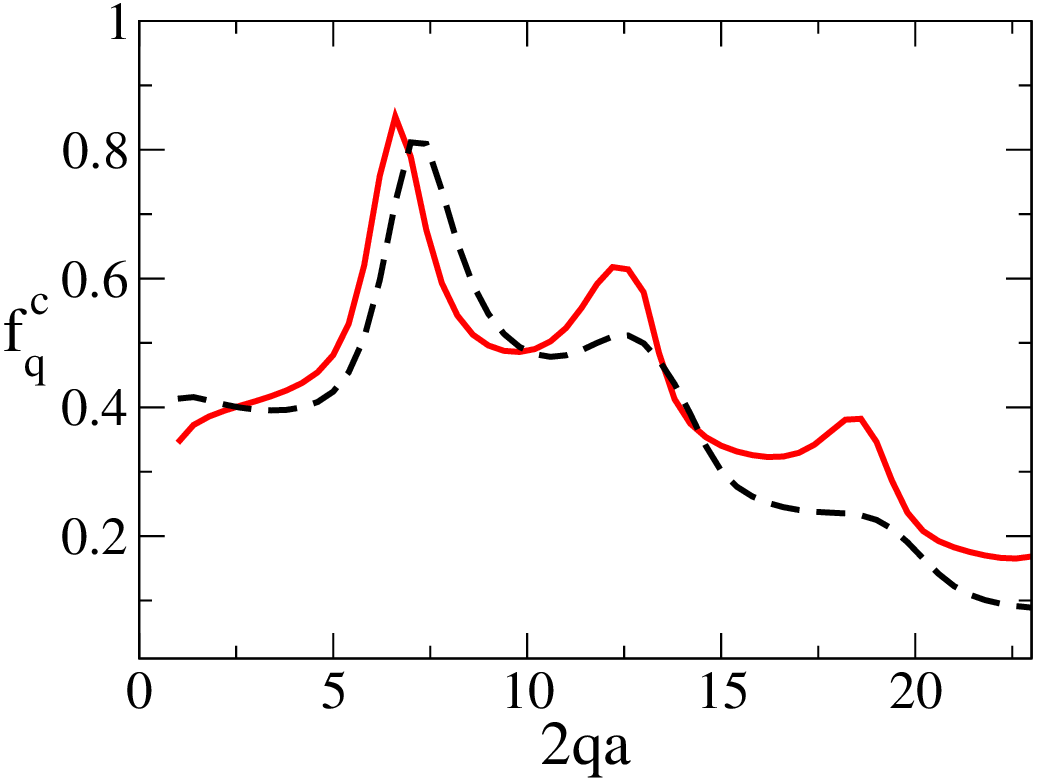}
  \caption{\label{fig:fq2D3Deps} Critical glass-form
    factors $f_q^c$ at the respective transition points for
    $\varepsilon = 0.5$ in 2D (full curve) and 3D (dashed curves).}
\end{figure}

It is seen in Fig.~\ref{fig:fq2Deps} that for smaller $\varepsilon$
the order parameter $f_q^c$ decays more slowly for large wave numbers
indicating a tighter localization. In comparison to 3D, the transition
in 2D exhibits individually sharper peaks and overall a tighter
localization for the same dissipation, cf. Fig.~\ref{fig:fq2D3Deps}.
It has been shown for data from simulation \cite{Sperl2003a} and
experiments in colloidal suspensions \cite{vanmegen95,bayer07}, that
the MCT predictions for the $f_q^c$ are typically accurate to around
20\%. Hence, rather than fitting individual $f_q^c$ directly to
measurements and numerical calculations, experimental and simulation
data can be expected to follow the distributions shown on the
20\%-level and exhibit trends with variation of $\varepsilon$ as
indicated here.

It is seen in Fig.~\ref{fig:coh05_42}, that the critical law can only
be observed without corrections for states closer than 0.1\% to the
transition point. Also, the von Schweidler law in
Fig.~\ref{fig:coh05_70} is only valid for an intermediate regime after
the plateau. The regimes of applicability for the asymptotic scaling
laws are therefore similar to the elastic case, and corrections to
scaling are expected to follow the known trends
\cite{franosch+fuchs97}.

\section{Tagged Particle Dynamics}
\label{sec:tagg-part-dynam}

The results of granular MCT for the tagged particle dynamics has been
discussed in II. Here, we will focus on the derivation of the MCT
equations.

The incoherent scattering function, $\phi_s(q,t)$, captures the tagged
particle dynamics. This includes the mean square displacement $\delta
r^2(t) = \avr{[\vec r_s - \vec r_s(t)]^2}$ which appears as an
expansion coefficient of the incoherent scattering function
$\phi_s(q,t) = 1 - q^2\delta r^2(t)/6 + \mathcal O(q^4)$ and the
diffusivity $6D_{\infty} = \lim_{t\to\infty}\delta r^2(t)/t$
\cite{boon+yip92}.

\subsection{Equation of Motion}
\label{sec:equation-motion}

Following the reasoning that one should account for the conserved
quantities and only for the conserved quantities explicitly, one would
assume that the equation of motion for a tagged particle should be
first order in time. The density $\rho_s$ being the only conserved
quantity as the momentum of the tagged particle is all but
conserved. It has been shown, though, that a consistent treatment of
the tagged particle dynamics in fact requires an equation of motion
which is second order in time \cite{cichocki+hess87,pitts+andersen00},
thus effectively reintroducing the tagged particle momentum as a
macroscopic observable.

We follow that reasoning and introduce the projector
\begin{equation}
  \label{eq:67}
  \P_s = \sum_{\vec q}\ket{\rho_{\vec q}^s}\bra{\rho_{\vec q}^s} +
  \sum_{\vec q}\ket{j_{\vec q}^{sL}}\bra{j_{\vec q}^{sL}}/T.
\end{equation}
Together with the microscopic state $\vec a_{\vec q}^s = (\rho_{\vec
  q}^s, j_{\vec q}^{sL}/\sqrt T)$, it yields an equation of motion for
$\phi_s(q,t)$, formally identical to Eq.~\eqref{eq:18},
\begin{equation}
  \label{eq:68}
  \begin{aligned}
    \ddot\phi_s(q,t) &+ \nu_q\dot\phi_s(q,t) + \Omega_s^s\phi_s(q,t)\\
    &+ \Omega_s^2\int_0^t\!d\tau\,m_s(q,t-\tau)\dot\phi_s(q,\tau)\\
    &+ \Omega_s^2\int_0^t\!d\tau\,L_s(q,t-\tau)\phi_s(q,\tau) = 0,
  \end{aligned}
\end{equation}
with $\dot\phi_s(q, t = 0) = 0$ and $\phi_s(q, t = 0) = 1$. Moreover,
$\Omega^s_{j\rho} = \avr{j_{\vec q}^{sL}|\Lv_+\rho_{\vec q}^s}/\sqrt T
= q\sqrt T$ and $\Omega^s_{\rho j} = q\sqrt T$ do not depend on the
coefficient of restitution as shown in
appendix~\ref{sec:freq-omeg-js}. Hence, $\Omega_s^2 :=
\Omega^s_{j\rho}\Omega^s_{\rho j} = q^2T$ is identical to the
corresponding quantity of the molecular fluid. This implies that
looking at the probability density of the tagged particle on
macroscopic time and length scales at very low densities, such that
the memory kernel can be neglected, the microscopically broken time
reversal symmetry is unobservable.

The memory kernels are given by
\begin{subequations}
\begin{align}
  \label{eq:69}
  m_s(q,t) &= \avr{F_{\vec q}^{s\dagger}\tilde{\mathsf{U}}(t)F_{\vec q}^s}/q^2T^2,\\
  L_s(q,t) &= \avr{J_{\vec q}^{s\dagger}\tilde{\mathsf{U}}(t)F_{\vec q}^s}/qT,
\end{align}
\end{subequations}
with the fluctuating forces $F_{\vec q}^s = \Q_s\Lv_+j_{\vec q}^{sL}$
and $F_{\vec q}^{s\dagger} = \Q_s\Lv_+^{\dagger}j_{\vec q}^{sL}$ and
the fluctuating current is $J_{\vec q}^{s\dagger} =
\Q_s\Lv_+^{\dagger}\rho_{\vec q}^s$.

\subsection{The Mode Coupling Approximation}
\label{sec:mode-coupl-appr-1}

We introduce a projection operator to describe the coupling between
the tagged particle and the host fluid
\begin{equation}
  \label{eq:70}
  \P_2^s = N\sum_{\vec k,\vec p}\ket{\rho_{\vec k}\rho^s_{\vec
      p}}\bra{\rho_{\vec k}\rho_{\vec p}^s}/S_k.
\end{equation}
The corresponding mode coupling approximation reads
\begin{equation}
  \label{eq:71}
  \begin{aligned}
    \tilde{\mathsf U}(t) &\approx \P_2^s\tilde{\mathsf U}(t)\P_2^s\\
    &\approx N\sum_{\vec k, \vec p}\ket{\rho_{\vec k}\rho^s_{\vec p}}
    \phi(k,t)\phi_s(p,t)\bra{\rho_{\vec k}\rho^s_{\vec p}}/S_k.
  \end{aligned}
\end{equation}

Within the MCA we find that $L_s(q,t) = 0$ and
\begin{multline}
  \label{eq:72}
  m_s[\phi,\phi_s](q,t) \approx\\
  \frac1{q^2T^2}\sum_{\vec k, \vec p}
  \mathcal V_{\vec q\vec k\vec p}^s\mathcal W_{\vec q\vec k\vec p}^s\phi(k,t)\phi_s(p,t),
\end{multline}
where
\begin{subequations}
\begin{align}
  \label{eq:73}
  \mathcal V_{\vec q\vec k\vec p}^s 
  &= \sqrt{N/S_k}\avr{j_{\vec q}^{sL}|\Lv_+\Q_s\rho_{\vec k}\rho_{\vec p}^s},\\
  \mathcal W_{\vec q\vec k\vec p}^s 
  &= \sqrt{N/S_k}\avr{\rho_{\vec k}\rho_{\vec p}^s|\Q_s\Lv_+j_{\vec q}^{sL}}.
\end{align}
\end{subequations}
Here,
\begin{subequations}
\begin{equation}
  \label{eq:74}
  \mathcal V^s_{\vec q\vec k\vec p} = \frac{T}{\sqrt{NS_k}}
  (\uvec q\cdot\vec k)(S_k - 1)\delta_{\vec q,\vec k+\vec p},
\end{equation}
is known from the literature
\cite{wahnstroem+sjoegren82,bengtzelius+goetze84}\footnote{We are not
  aware of a published derivation within the projection operator
  formalism, though. We present it in
  appendix~\ref{sec:vertex-mathcal-v_vec}} and one finds
(cf. appendix~\ref{sec:vertex-mathcal-w_vec})
\begin{equation}
  \label{eq:75}
  \mathcal W^s_{\vec q\vec k\vec p} = \frac{1+\varepsilon}{2}\frac{T}{\sqrt{NS_k}}
  (\uvec q\cdot\vec k)(S_k - 1)\delta_{\vec q,\vec k+\vec p}.
\end{equation}
\end{subequations}
For the vertices, the loss of detailed balance reappears also for the
tagged particle dynamics. 

Together, Eqs.~\eqref{eq:74} and \eqref{eq:75} yield
\begin{multline}
%  \label{eq:76}
  m_s[\phi,\phi_s](q,t)\approx\\
  \frac{1+\varepsilon}{2}\frac{n}{q^2}\!
  \int d^D\!k\,S_k(\uvec q\cdot\vec k)^2c_k^2
  \phi(k, t)\phi_s(|\vec q-\vec k|,t).\notag
\end{multline}

\subsection{The approximate equation of motion}
\label{sec:appr-equat-moti-1}

Finally, the equation of motion for the incoherent scattering
function reads
\begin{multline}
  \label{eq:78}
  \ddot\phi_s(q,t) + \nu_q\dot\phi_s(q,t) + \Omega_s^2\phi_s(q,t)\\
  + \Omega_s^2\int_0^{\infty}\!d\tau\, m_s[\phi,\phi_s](q,t-\tau)
  \dot\phi_s(q,\tau) = 0,
\end{multline}
capturing the coupling of the tagged particle dynamics to the dynamics
of the host fluid as reported in II.

The tagged particle is enslaved to the host fluid on macroscopic time
scales. Consequently, it also develops persistent correlations, $f_q^s
= \lim_{t\to\infty}\phi_s(q,t)$, at the critical density
$\varphi_c$. They can be calculated from the equation
\begin{equation}
  \label{eq:86}
  \frac{f_q^s}{1 - f_q^s} = m[f, f^s](q).
\end{equation}
A detailed discussion of the solutions of Eqs.~(\ref{eq:78},\ref{eq:86}) was
given in II.

\section{Discussion}
\label{sec:discussion}

The granular MCT, which includes and extends MCT for elastic hard
spheres, shows that the dynamics of a driven granular fluid is for one
remarkably similar to the equilibrium dynamics and at the same time
fundamentally different. It is similar in that there is always a glass
transition, accompanied by the two step relaxation scenario of dynamic
correlation functions and diverging time scales. As both the order
parameter $f_q(\varepsilon)$ and especially the critical exponents
$a(\varepsilon)$ and $b(\varepsilon)$ depend on the coefficient of
restitution $\varepsilon$, already slightly dissipative interactions
($\varepsilon \lesssim 1$) destroy the universality of the dynamics on
long time scales, which is observed in elastic systems with either
Newtonian or Brownian dynamics
\cite[I,][]{Loewen1991,Gleim1998}. The change of $\varepsilon$ will
be detectable even on macroscopic time scales, in particular by
observing the exponent $b(\varepsilon)$. This shows that the
combination of dissipative collisions and driving cannot be mapped to
an effective elastic hard sphere system with an effective temperature
$T_{\mathrm{eff}}$, conceivably different from the granular
temperature, $T$. Such a mapping, would allow to find a scaling
function such that $\phi(q,t;\varepsilon) = 
\tilde\phi(q,t/t_0(\varepsilon))$.

The phase diagram in the $(T,\varphi)$-plane is still an open problem.
The jamming density is
defined for athermal ($T=0$) systems, while the granular glass
transition is independent of temperature but assumes a finite
temperature $T > 0$ to sustain a fluid phase.
It is not obvious, if and how they are connected.
Our results suggest that the glass transition density, $\varphi_c$, is
always strictly smaller than the quasi static jamming density
$\varphi_J$. However, MCT is known to underestimate $\varphi_c$. 
What happens for densities $\varphi \ge \varphi_c(\varepsilon)$ larger
than the critical density?  At the glass transition, the
$\alpha$-relaxation rate $\tau^{-1}$ diverges. Consequently, in every
compression protocol using a small but finite compression rate, at
some density the compression rate will be larger than the
$\alpha$-relaxation rate $\tau^{-1}$.  From then on, the evolution of
the system will be restricted to a subset of phase space. The
packings, which are reached from that subset by further compression
will also be restricted to a subset of all packings and that might not
even include those of highest density. Even if the ideal glass transition 
is destroyed by
processes which are ignored within MCT, the enormous increase of
relaxation times will be prohibitive for all practical purpose.

Apart from the Enskog term, Eqs.~(\ref{eq:32a},\ref{eq:32b}), the
equations of motion are formally identical in two and three space
dimensions. Hence the glass transition is qualitatively similar in two
and three dimensions, with however different values for the critical
density and the critical exponents. Compared to $D=3$, the glass from
factors, $f_q$, decay slower in reciprocal space for $D=2$, indicating a
stronger localization in two space dimensions.

In the final equations of motion, Eqs.~\eqref{eq:30} and \eqref{eq:78}, 
the driving force $\vec\xi$ appears only implicitly. While driving is 
crucial to achieve a stationary state, beyond that it does not alter the 
relaxation rates $\Omega_{ab},\Omega_{ab}^s$ and does not enter into the 
couplings to densities. Driving contributions would appear in the linear 
theory, if the (kinetic) energy, i.e., the granular temperature was 
included as a dynamic field. However this is hard to justify in a granular 
fluid, where kinetic energy is dissipated locally and hence not a 
hydrodynamic variable. In terms of the MCT, a coupling to the currents in 
$\P_2$ [Eq.~\eqref{eq:22}] would include explicit driving terms. Such a 
coupling was considered in the original mode coupling approaches 
\cite{mazenko74,sjoegren80b}, but it's relevance even in equilibrium 
fluids remains unclear.

\section{Conclusion and Outlook}
\label{sec:conclusion}

We considered randomly driven inelastic smooth hard disks (in $D=2$)
and spheres (in $D=3$). We systematically derived equations of motion
for the coherent scattering function, $\phi(q, t)$, and the incoherent
scattering function, $\phi_s(q,t)$.

The equations of motion are formally identical to the ones for elastic
hard sphere or disk fluids in thermal equilibrium but acquire a
nontrivial dependence on the coefficient of restitution, $\varepsilon$.
A transition to a glassy state, indicated by a nonzero value of the
order parameter, $f_q$, appears through the bifurcation scenario of
mode coupling theory.  Like in thermal equilibrium, the spatial
dimension of the system only enters via the static structure
factors. In both dimensions, the critical packing fraction increases
the more dissipative the particles are.

The dynamics around the plateau in the scattering functions is
described by power laws with exponents, that are functions of the
coefficient of restitution, $\varepsilon$. Together with the
$\varepsilon$-dependence of the order parameter, $f_q$, this shows
that the dynamics fundamentally changes upon varying the coefficient
of restitution. In contrast, the difference between Newtonian and
Brownian dynamics in thermal equilibrium, can be absorbed in the
redefinition of the microscopic time scale. Also, a reduced long 
wavelength speed of sound is predicted for granular fluids.

One can hardly expect to observe a glass transition in a fluid of
monodisperse hard spheres, because the system would quickly
crystallize. To slow down crystal nucleation, usually binary mixtures
with a small size difference are used
\cite{zaccarelli+valeriani09}. For fluids in thermal equilibrium, it
was found that a MCT for mixtures does not yield results that differ
drastically from those for the monodisperse idealization
\cite{barrat+latz90}. For mixtures of granular particles, a new
complication will be the non-equipartition of energy between the
mixture species \cite{barrat+trizac02,uecker+kranz09}.

So far we only derived equations for the correlation functions of
spontaneous fluctuations. In a fluid in thermal equilibrium, this
immediately entails knowledge about the corresponding response
functions via the fluctuation dissipation theorem (FDT)
\cite{degroot+mazur}. In fact, a lot of the experimental measurements
are concerned with response spectra \cite{jaeckle86}. The existence
and form of a generalized FDT in driven granular fluids and more
generally in systems far from equilibrium is a subject of active
research \cite{marconi+puglisi08}.

It would certainly be desirable to weaken the assumptions made on the
stationary phase space distribution function $\varrho(\Gamma)$. So
far, we ignore correlations between the velocities of different
particles which are known to be present in driven granular fluids
\cite{pagonabarraga+trizac01}. In light of the fact, that the single
particle velocity distribution function is well represented by a
simple Gaussian, these correlations can presumably be neglected as a
first approximation. More serious are the static correlations, such as
$S(q)$, which are known from simulations to differ from their elastic
counterparts used here. However, we can easily incorporate data for
the simulated structure factors into our approach; work along these
lines is in progress.

The results presented above deal with a specific, highly idealized
system. It is a natural question to ask how robust these results are
qualitatively. No qualitative changes are expected for a speed
dependent coefficient of restitution $\varepsilon =
\varepsilon(v)$. Also models that can be described by an effective
coefficient of restitution \cite{schaefer+dippel96} like, e.g., the
spring-dashpot model are expected to show a nonequilibrium glass
transition. The inclusion of inter-particle friction or the treatment
of different driving forces will likely pose a number of
challenges. Such changes might lead to equations of motion and results
qualitatively different from the ones discussed here.

\begin{acknowledgments}
We acknowledge financial support by the DFG (FG1394). 
\end{acknowledgments}

\appendix

\section{The Granular Yvon-Born-Green Relation}
\label{sec:granular-born-green}

The Yvon-Born-Green (YBG) relation between the pair- and the triplet correlation
function follows from the identity
\begin{equation}
  \label{eq:48}
  \begin{aligned}
    \nabla_1g(r_{12}) &= (\varrho^{-1}\nabla_{12}\varrho)g(r_{12})\\
    &+ n\int d^Dr_3\,g_3(\vec r_1,\vec r_2,\vec r_3)(\varrho^{-1}\nabla_{13}\varrho),
  \end{aligned}
\end{equation}
where $\varrho^{-1}$ is the pseudo-inverse of the distribution
function \cite{born+green46}. For elastic hard spheres, where
$\varrho(\Gamma) \propto \prod_{i<j}\Theta(r_{ij} - 2a)$ This yields
the known relation \cite{hansen+mcdonald06}
\begin{equation}
  \label{eq:62}
  \begin{aligned}
    \nabla_1g(r_{12}) &= \uvec r_{12}\delta(r_{12} - 2a)g(r_{12})\\
    &+ n\int d^Dr_3\,\uvec r_{13}\delta(r_{13} - 2a)g_3(\vec r_1,\vec r_2,\vec r_3).
  \end{aligned}
\end{equation}
For the inelastic hard spheres, there must be an additional spatial
dependence of the distribution function, depending on the coefficient
of restitution, $\varepsilon$, or otherwise, e.g., the structure factor
$S_q$ would not depend on $\varepsilon$. 

Overlapping configurations still have zero probability and because of
homogeneity, only relative distances play a role. Therefor, the
distribution function will be of the form $\varrho(\Gamma)\propto
\prod_{i<j}\Theta(r_{ij} - 2a)\vartheta_{\varepsilon}(r_{ij})$ with a
unknown function $\vartheta_{\varepsilon}(r) > 0$. With this, we get a
granular hard sphere YBG relation
\begin{equation}
  \label{eq:65}
  \begin{aligned}
    \nabla_1g(r_{12}) &= \uvec r_{12}\delta(r_{12} - 2a)g(r_{12})
    + g(r_{12})\nabla_1\ln\vartheta_{\varepsilon}(r_{12})\\
    &+ n\int d^Dr_3\,\uvec r_{13}\delta(r_{13} - 2a)g_3(\vec r_1,\vec r_2,\vec r_3)\\
    &+ n\int d^Dr_3\,g_3(\vec r_1,\vec r_2,\vec r_3)
    \nabla_1\ln\vartheta_{\varepsilon}(r_{13}).
  \end{aligned}\notag
\end{equation}

Unfortunately, virtually nothing is known about the function
$\vartheta_{\varepsilon}(r)$. Therefor we use the elastic hard sphere
YBG relation which means we make the nontrivial approximation
\begin{multline}
  \label{eq:66}
  g(r_{12})\nabla_1\ln\vartheta_{\varepsilon}(r_{12})\\
  \approx -n\int d^Dr_3\,g_3(\vec r_1,\vec r_2,\vec r_3)
    \nabla_1\ln\vartheta_{\varepsilon}(r_{13})
\end{multline}
which may be more general than setting
$\vartheta_{\varepsilon}(r)\equiv1$.

On the next level, we have
\begin{equation}
  \label{eq:87}
  \begin{aligned}
    \nabla_1g_3(123) 
    &= [\uvec r_{12}\delta(r_{12} - 2a) + \uvec r_{13}\delta(r_{13} - 2a)]
    g_3(123)\\
    &+ g_3(123)
    \nabla_1[\ln\vartheta_{\varepsilon}(r_{12}) + \ln\vartheta_{\varepsilon}(r_{13})]\\
    &+ n\int d^Dr_4\,\uvec r_{14}\delta(r_{14} - 2a)g_4(1234)\\
    &+ n\int d^Dr_4\,g_4(1234)\nabla_1\ln\vartheta_{\varepsilon}(r_{14})
  \end{aligned}\notag
\end{equation}
using the abbreviation $i\equiv\vec r_i$. In Eq.~\eqref{eq:51} below, we
use the approximation
\begin{multline}
  \label{eq:88}
  g_3(123)
  \nabla_1[\ln\vartheta_{\varepsilon}(r_{12}) + \ln\vartheta_{\varepsilon}(r_{13})]\\
  \approx -n\int d^Dr_4\,g_4(1234)\nabla_1\ln\vartheta_{\varepsilon}(r_{14})
\end{multline}

\section{Matrix Elements}
\label{sec:matrix-elements}

\subsection{The Frequency $\Omega_{jj}$ in Two Dimensions}
\label{sec:freq-omeg-two}

We have to determine
\begin{equation}
  \label{eq:B1}
  \avr{j_{\vec q}^L|\Lv_+j_{\vec q}^L} 
  = i\frac{N(N-1)}{2}\avr{j_{\vec q}^L|\mathcal T_{12}^+j_{\vec q}^L}
\end{equation}
where all other contributions vanish due to parity. Explicitly, this
reads
\begin{widetext}
  \begin{equation}
    \label{eq:B2}
    \avr{j_{\vec q}^L|\Lv_+j_{\vec q}^L} 
    = \frac{1+\varepsilon}{2}i\avr{(\uvec q\cdot\vec v_1)(\uvec
      q\cdot\uvec r_{12})(\uvec r_{12}\cdot\vec v_{12})^2\Theta(-\uvec
      r_{12}\cdot\vec v_{12})\delta(r_{12} - 2a)\big(e^{i\vec q\cdot\vec r_{12}}-1\big)},
  \end{equation}
  where the three particle term vanishes, again, due to
  parity. Introducing the relative velocity $\vec v := (\vec v_1 -
  \vec v_2)/\sqrt2$, the velocity averages can be evaluated
  \begin{equation}
    %\label{eq:B3}
    \begin{aligned}
      \avr{(\uvec q\cdot\vec v_1)(\uvec r_{12}\cdot\vec
        v_{12})^2\Theta(-\uvec r_{12}\cdot\vec v_{12})}
      &= \sqrt2(\uvec q\cdot\uvec r_{12})
      \avr{(\uvec r_{12}\cdot\vec v_{12})^3\Theta(-\uvec r_{12}\cdot\vec v_{12})}\\
      &= \frac{\sqrt2}{2\pi T}(\uvec q\cdot\uvec r_{12})
      \int_0^{\infty}\!dv\int_{\pi/2}^{3\pi/2}\!d\varphi\,v^4\cos^3\varphi e^{-v^2/2T}\\
      &= -2T\sqrt{T/\pi}(\uvec q\cdot\uvec r_{12}).
    \end{aligned}
  \end{equation}
\end{widetext}
The remaining spatial average reads
\begin{multline}
  \label{eq:B4}
  \avr{(\uvec q\cdot\uvec r_{12})^2\delta(r_{12}-2a)\big(e^{i\vec
      q\cdot\vec r_{12}}-1\big)}\\
  = -\frac{2a\chi}{V}\int_0^{2\pi}d\varphi\,\cos^2\varphi
  \big(1 - e^{2iaq\cos\varphi}\big).
\end{multline}

One finds 
\begin{equation}
  \label{eq:B5}
  \frac1\pi\int_0^{2\pi}d\varphi\,\cos^2\varphi e^{iz\cos\varphi}
  = -2\frac{d^2J_0(z)}{dz^2},
\end{equation}
i.e.,
\begin{multline}
  \label{eq:B6}
  \avr{(\uvec q\cdot\uvec r_{12})^2\delta(r_{12}-2a)\big(e^{i\vec
      q\cdot\vec r_{12}}-1\big)}\\
  = -\frac{2\pi na\chi}{N}[1 + 2J_0''(2aq)].
\end{multline}
Collecting terms, one arrives at Eq.~\eqref{eq:32a}.

\subsection{The Frequency $\Omega_{\rho j}$}
\label{sec:freq-omeg-j}

The driving contribution vanishes and the free streaming contribution
yields
\begin{equation}
  \label{eq:40}
  \begin{aligned}
    \avr{\rho_{\vec q}|\Lv_0j_{\vec q}^L} 
    &= \frac{q}{N^2}\avr{\sum_{j,k}(\uvec q\cdot\vec v_k)^2e^{-i\vec
        q\cdot\vec r_{jk}}}\\
    &= qTS_q/N.
  \end{aligned}
\end{equation}
The collisional contribution reads
\begin{widetext}
  \begin{equation}
    \label{eq:41}
    \frac{N(N-1)}{2}\avr{\rho_{\vec q}|\mathcal T_{12}j_{\vec q}^L}
    = \frac{1+\varepsilon}{4}\uvec q\cdot
    \avr{(\uvec r_{12}\cdot\vec v_{12})^2\uvec r_{12}\Theta(-\uvec
      r_{12}\cdot\vec v_{12})\delta(r_{12} - 2a)\big(e^{i\vec
        q\cdot\vec r_2} - e^{i\vec q\cdot\vec r_1}\big)\sum\nolimits_je^{-i\vec
        q\cdot\vec r_j}}.
  \end{equation}
\end{widetext}
The velocity integration yields a factor $T/2$ while the spatial
average can be rewritten as
\begin{multline}
  \label{eq:42}
  \avr{\uvec r_{12}\delta(r_{12} - 2a)\big(
    e^{i\vec q\cdot\vec r_2} - e^{i\vec q\cdot\vec r_1}
    \big)\sum\nolimits_je^{-i\vec q\cdot\vec r_j}}\\
  = 2\avr{\uvec r_{12}\delta(r_{12} - 2a)e^{-i\vec q\cdot\vec r_{12}}}\\
  + 2(N-2)\avr{\uvec r_{12}\delta(r_{12} - 2a)e^{-i\vec q\cdot\vec
      r_3}e^{i\vec q\cdot\vec r_2}}.
\end{multline}
Application of the \textsc{YBG} relation to the second term yields
\begin{multline}
  \label{eq:43}
  N\avr{\uvec r_{12}\delta(r_{12} - 2a)e^{-i\vec q\cdot\vec
      r_3}e^{i\vec q\cdot\vec r_2}}\\
  = -\frac1N(S_q-1) - \avr{\uvec r_{12}\delta(r_{12} - 2a)e^{-i\vec
      q\cdot\vec r_{12}}}, 
\end{multline}
i.e., the second term in Eq.~\eqref{eq:43} cancels the first term in
Eq.~\eqref{eq:42}. Combining the remaining terms we arrive at
Eq.~\eqref{eq:15}. 

\subsection{The Vertex $\mathcal W_{\vec q\vec k\vec p}$}
\label{sec:vertex-mathal-w_vec}

As the vertex is linear in $\vec v_i$, there is no contribution from
the driving, $i\Lv_D^+$. Expanding the projector $\Q_c$, the vertex
reads
\begin{multline}
  \label{eq:31}
  \mathcal W_{\vec q\vec k\vec p} 
  = N\avr{\rho_{\vec k}\rho_{\vec p}|\Lv_+j_{\vec q}^L}/S_p\\
  - N^2\avr{\rho_{\vec k}\rho_{\vec p}|\rho_{\vec q}}
  \avr{\rho_{\vec q}|\Lv_+j_{\vec q}^L}/S_pS_q.
\end{multline}
The free streaming contribution to the first term is given by
\begin{equation}
  \label{eq:32}
  \begin{aligned}
  \avr{\rho_{\vec k}\rho_{\vec p}|\Lv_0j_{\vec q}^L} 
  &= qT\avr{\rho_{\vec k}\rho_{\vec p}|\rho_{\vec q}}\\
  &= \frac{qT}{N^2}\delta_{\vec k + \vec p, \vec q}S^{(3)}(\vec k, \vec p).
\end{aligned}
\end{equation}

For the collisional contribution we find
\begin{widetext}
  \begin{equation}
    \label{eq:33}
    \frac{N(N-1)}{2}\avr{\rho_{\vec k}\rho_{\vec p}|\mathcal T_{12}^+j_{\vec q}^L}
    = i\frac{1+\varepsilon}{4N}T\uvec q\cdot\avr{\sum_{j,k}e^{-i\vec
        k\cdot\vec r_j}e^{-i\vec p\cdot\vec r_k}\uvec r_{12}\delta(r_{12}-2a)
      \big(e^{i\vec q\cdot\vec r_2} - e^{i\vec q\cdot\vec r_1}\big)}
  \end{equation}
\end{widetext}
The average on the right hand side shall be abbreviated as
$\avr{jk|12}$. Then this can be expanded as
\begin{multline}
  \label{eq:34}
  \avr{jk|12} = \avr{11|12} + \avr{22|12} + \avr{12|12} + \avr{21|12}\\
  + N(\avr{13|12} + \avr{23|12} + \avr{31|12} + \avr{32|12})\\ 
  + N\avr{33|12} + N^2\avr{34|12}
\end{multline}
Exploiting the symmetries, this can be simplified to
\begin{multline}
  \label{eq:35}
  \avr{jk|12} = 2\avr{11|12} + \avr{12|12} + 2N\avr{13|12}\\
  + N\avr{33|12} + N^2\avr{34|12}
\end{multline}
Proceeding term by term we first find
\begin{equation}
  \label{eq:36}
  \avr{11|12} = \avr{e^{i(\vec k+\vec p-\vec q)\cdot\vec r_1}\uvec
    r_{12}\delta(r_{12} - 2a)\big(e^{i\vec q\cdot\vec r_{12}} - 1\big)}\notag
\end{equation}
which can be reduced to
\begin{equation}
  \label{eq:37}
  \begin{aligned}
    \avr{11|12} &= \delta_{\vec k+\vec p,\vec q}
    \avr{\uvec r_{12}\delta(r_{12} - 2a)e^{i\vec q\cdot\vec r_{12}}}\\
    &\equiv \delta_{\vec k+\vec p,\vec q}\vec G(\vec q).
  \end{aligned}
\end{equation}
The second term can be reduced to an equivalent expression
\begin{equation}
  \label{eq:38}
  \begin{aligned}
    \avr{12|12} &= \avr{e^{i(\vec k+\vec p-\vec q)\cdot\vec r_1}\uvec
      r_{12}\delta(r_{12} - 2a)e^{i\vec q\cdot\vec r_{12}}}\\
    &- \avr{e^{i(\vec k+\vec p-\vec q)\cdot\vec r_1}\uvec
      r_{12}\delta(r_{12} - 2a)e^{i(\vec k-\vec q)\cdot\vec r_{12}}},
  \end{aligned}\notag
\end{equation}
i.e.,
\begin{equation}
  \label{eq:39}
  \avr{12|12} = \delta_{\vec k+\vec p,\vec q}[\vec G(\vec k) + \vec G(\vec p)].
\end{equation}

The first three particle term
\begin{equation}
  \label{eq:44}
  \begin{aligned}
    \avr{13|12} &= \avr{e^{i(\vec k + \vec p - \vec q)\cdot\vec
        r_2}e^{i\vec k\cdot\vec r_{12}}e^{i\vec p\cdot\vec
        r_{32}}\uvec r_{12}\delta(r_{12} - 2a)}\\
    &- \avr{e^{i(\vec k - \vec q)\cdot\vec r_1}e^{i\vec p\cdot\vec
        r_3}\uvec r_{12}\delta(r_{12}- 2a)}
  \end{aligned}\notag
\end{equation}
requires a little more work. The first term shall be abbreviated as
\begin{multline}
  \label{eq:45}
  \avr{e^{i(\vec k + \vec p - \vec q)\cdot\vec
      r_2}e^{i\vec k\cdot\vec r_{12}}e^{i\vec p\cdot\vec
      r_{32}}\uvec r_{12}\delta(r_{12} - 2a)}\\
  = \delta_{\vec k+\vec p,\vec q}\vec H(\vec k,\vec p).
\end{multline}
The second term can be simplified with the help of the \textsc{YBG}
relation
\begin{multline}
  \label{eq:46}
  \avr{e^{i(\vec k - \vec q)\cdot\vec r_1}e^{i\vec p\cdot\vec
      r_3}\uvec r_{12}\delta(r_{12}- 2a)}\\
  = \frac{1}{N^2}\delta_{\vec k+\vec p,\vec q}
  [i\vec p(S_p-1) + N\vec G(\vec p)].
\end{multline}

Similarly, the second three particle term 
\begin{equation}
  \label{eq:47}
  \avr{33|12}
  = \avr{e^{i(\vec k + \vec p)\cdot\vec r_3}\uvec
    r_{12}\delta(r_{12} - 2a)\big(e^{-i\vec q\cdot\vec r_2} -
    e^{-i\vec q\cdot\vec r_1}\big)}\notag
\end{equation}
can be reduced by employing the \textsc{YBG} relation
\begin{equation}
  \label{eq:49}
  \avr{33|12} = -\frac{2}{N^2}\delta_{\vec k + \vec p, \vec q}
  [i\vec q(S_q-1) + N\vec G(\vec q)].
\end{equation}

The four particle term
\begin{equation}
  \label{eq:50}
  \avr{34|12} = \avr{e^{i\vec k\cdot\vec r_3}e^{i\vec p\cdot\vec
      r_4}\uvec r_{12}\delta(r_{12}-2a)\big(e^{-i\vec q\cdot\vec r_2}
    - e^{-i\vec q\cdot\vec r_1}\big)}\notag
\end{equation}
is naturally the most involved. Using the higher order \textsc{YBG}
relation it reads
\begin{widetext}
  \begin{equation}
    \label{eq:51}
    \begin{aligned}
      \avr{34|12} = &-\frac{2}{NV^3}\int d^Dr_2d^Dr_3d^Dr_4\,
      e^{-i\vec q\cdot\vec r_2}e^{i\vec k\cdot\vec r_3}e^{i\vec
        p\cdot\vec r_4}\frac{\partial}{\partial\vec r_2}
      g_3(\vec r_2,\vec r_3,\vec r_4)\\
      &+ \frac{2}{NV^3}\int d^Dr_2d^Dr_3d^Dr_4\,
      e^{-i\vec q\cdot\vec r_2}e^{i\vec k\cdot\vec r_3}e^{i\vec
        p\cdot\vec r_4}\uvec r_{23}\delta(r_{23} - 2a)
      g_3(\vec r_2,\vec r_3,\vec r_4)\\
      &+ \frac{2}{NV^3}\int d^Dr_2d^Dr_3d^Dr_4\,
      e^{-i\vec q\cdot\vec r_2}e^{i\vec k\cdot\vec r_3}e^{i\vec
        p\cdot\vec r_4}\uvec r_{24}\delta(r_{24} - 2a)
      g_3(\vec r_2,\vec r_3,\vec r_4).
    \end{aligned}
  \end{equation}
  Partial integration in the first term and the extraction of the
  momentum conservation constraint yields
  \begin{equation}
    \label{eq:52}
    \begin{aligned}
      \avr{34|12} = &-\frac{2i\vec q}{NV^2}\delta_{\vec k + \vec p, \vec q}
      \int d^Dr_{23}d^Dr_{24}\,e^{-i\vec k\cdot\vec r_{23}}e^{-i\vec p\cdot\vec r_{24}}
      g_3(\vec r_{23}, \vec r_{24})\\
      &+\frac{2}{NV^2}\delta_{\vec k + \vec p, \vec q}
      \int d^Dr_{23}d^Dr_{24}\,e^{-i\vec k\cdot\vec r_{23}}e^{-i\vec p\cdot\vec r_{24}}
      g_3(\vec r_{23}, \vec r_{24})\uvec r_{23}\delta(r_{23} - 2a)\\
      &+\frac{2}{NV^2}\delta_{\vec k + \vec p, \vec q}
      \int d^Dr_{23}d^Dr_{24}\,e^{-i\vec k\cdot\vec r_{23}}e^{-i\vec p\cdot\vec r_{24}}
      g_3(\vec r_{23}, \vec r_{24})\uvec r_{24}\delta(r_{24} - 2a).
    \end{aligned}
  \end{equation}
\end{widetext}
This leaves us with the relatively simple expression
\begin{equation}
  \label{eq:53}
  \begin{aligned}
    \avr{34|12} = &-\frac{2i\vec q}{N^3}\delta_{\vec k + \vec p, \vec q}
    [S^{(3)}(\vec k, \vec p) - S_k - S_p - S_q + 2]\\
    &- \frac2N\delta_{\vec k + \vec p, \vec q}
    [\vec H(\vec k, \vec p) + \vec H(\vec p, \vec k)].
  \end{aligned}\notag
\end{equation}

Most terms cancel to yield
\begin{equation}
  \label{eq:54}
  \avr{jk|12} = \frac{2i}{N}\delta_{\vec k + \vec p, \vec q}
  [\vec kS_p + \vec pS_k - \vec qS^{(3)}(\vec k, \vec p)]
\end{equation}
or
\begin{multline}
  \label{eq:55}
  \frac{N(N-1)}{2}\avr{\rho_{\vec k}\rho_{\vec p}|\mathcal T_{12}^+j_{\vec q}^L}\\
  = -\frac{1+\varepsilon}{2}\frac{T}{N^2}
  [(\uvec q\cdot\vec k)S_p + (\uvec q\cdot\vec p)S_k - qS^{(3)}(\vec k, \vec p)].
\end{multline}
Inserting Eq.~\eqref{eq:32} and Eq.~\eqref{eq:55} into
Eq.~\eqref{eq:31} and applying the convolution approximation yields
Eq.~\eqref{eq:17}.

\subsection{The Frequency $\Omega_{\rho j}^s$}
\label{sec:freq-omeg-js}

The free streaming contribution reads
\begin{equation}
  \label{eq:56}
  \avr{\rho_{\vec q}^s|\Lv_0j_{\vec q}^{sL}} 
  = q\avr{(\uvec q\cdot\vec v_s)^2} = qT
\end{equation}
and the collisional contribution
\begin{widetext}
  \begin{equation}
    \label{eq:57}
    (N-1)\avr{\rho_{\vec q}^s\mathcal T_{1s}^+j_{\vec q}^{sL}}
    = i\frac{1+\varepsilon}{2}N\avr{(\uvec r_{1s}\cdot\vec v_{1s})^2(\uvec
      q\cdot\uvec r_{1s})\Theta(-\uvec r_{1s}\cdot\vec
      v_{1s})\delta(r_{1s}-2a)} = 0
  \end{equation}
\end{widetext}
vanishes due to symmetry.

\subsection{The Vertex $\mathcal V_{\vec q\vec k\vec p}^s$}
\label{sec:vertex-mathcal-v_vec}

The left incoherent vertex is given as
\begin{equation}
  \label{eq:89}
  \mathcal V_{\vec q\vec k\vec p}^s 
  = \avr{j_{\vec q}^{sL}|\Lv_+\rho_{\vec k}\rho_{\vec p}^s}
  - \avr{j_{\vec q}^{sL}|\Lv_+\rho_{\vec q}^s}
  \avr{\rho_{\vec q}^s|\rho_{\vec k}\rho_{\vec p}^s}.
\end{equation}
The triple density correlator,
\begin{equation}
  \label{eq:59}
  \avr{\rho_{\vec q}^s|\rho_{\vec k}\rho_{\vec p}^s}
  = \frac1N\delta_{\vec k + \vec p, \vec q}S_k,
\end{equation}
is related to the structure factor. Moreover, we have
\begin{equation}
  \label{eq:90}
  \avr{j_{\vec q}^{sL}|\Lv_+\rho_{\vec k}\rho_{\vec p}^s}
  = k\avr{j_{\vec q}^{sL}|j_{\vec k}^L\rho_{\vec p}^s}
  + p\avr{j_{\vec q}^{sL}|\rho_{\vec k}j_{\vec p}^{sL}}
\end{equation}
as only the free streaming operator $i\Lv_0$ applies. The velocity
integration yield a factor of T
\begin{equation}
  \label{eq:91}
  \begin{aligned}
    \avr{j_{\vec q}^{sL}|\Lv_+\rho_{\vec k}\rho_{\vec p}^s}
    &= \frac{kT}{N}\avr{\rho_{\vec q}^s|\rho_{\vec k}^s\rho_{\vec p}^s}
    + pT\avr{\rho_{\vec q}^s|\rho_{\vec k}\rho_{\vec p}^s}\\
    &= \frac1N[(\uvec q\cdot\vec k)T + (\uvec q\cdot\vec p)S_k]
    \delta_{\vec k + \vec p, \vec q}.
  \end{aligned}
\end{equation}
Collecting terms one arrives at Eq.~\eqref{eq:74}.

\subsection{The Vertex $\mathcal W_{\vec q\vec k\vec p}^s$}
\label{sec:vertex-mathcal-w_vec}

The incoherent vertex is given as
\begin{equation}
  \label{eq:58}
  \mathcal W_{\vec q\vec k\vec p}^s = \avr{\rho_{\vec k}\rho_{\vec
      p}^s|\Lv_+j_{\vec q}^{sL}} - \avr{\rho_{\vec k}\rho_{\vec
      p}^s|\rho_{\vec q}^s}\avr{\rho_{\vec q}^s|\Lv_+j_{\vec q}^{sL}}.
\end{equation}

The free streaming contribution is simple 
\begin{equation}
  \label{eq:60}
  \avr{\rho_{\vec k}\rho_{\vec p}^s|\Lv_0j_{\vec q}^{sL}}
  = qT\avr{\rho_{\vec k}\rho_{\vec p}^s|\rho_{\vec q}^s} 
  = \frac{qT}{N}\delta_{\vec k + \vec p, \vec q}S_k.
\end{equation}

For the collisional part one finds with the velocity integration being
already performed
\begin{widetext}
  \begin{equation}
    \label{eq:61}
    \begin{aligned}
      (N-1)\avr{\rho_{\vec k}\rho_{\vec p}^s|\mathcal T_{1s}^+j_{\vec q}^{sL}}
      &= i\frac{1+\varepsilon}{2}\frac TN\avr{\sum\nolimits_je^{-i\vec
          k\cdot\vec r_j}e^{-i(\vec p - \vec q)\cdot\vec r_s}(\uvec
        q\cdot\vec r_{1s})\delta(r_{1s} - 2a)}\\
      &= i\frac{1+\varepsilon}{2}\frac TN\delta_{\vec k + \vec p - \vec q}
      \avr{\sum\nolimits_je^{-i\vec k\cdot\vec r_{js}}
        (\uvec q\cdot\vec r_{1s})\delta(r_{1s} - 2a)}.
    \end{aligned}
  \end{equation}
  The spatial average,
  \begin{equation}
    \label{eq:63}
    \avr{\sum\nolimits_je^{-i\vec k\cdot\vec r_{js}}
      (\uvec q\cdot\vec r_{1s})\delta(r_{1s} - 2a)}
    = \avr{e^{-i\vec k\cdot\vec r_{1s}}
      (\uvec q\cdot\vec r_{1s})\delta(r_{1s} - 2a)}
    +  N\avr{e^{-i\vec k\cdot\vec r_{2s}}
      (\uvec q\cdot\vec r_{1s})\delta(r_{1s} - 2a)},
  \end{equation}
\end{widetext}
can again be evaluated with the help of the \textsc{YBG}
relation. Applying it to the second term cancels the first term and we
get
\begin{multline}
  \label{eq:64}
  (N-1)\avr{\rho_{\vec k}\rho_{\vec p}^s|\mathcal T_{1s}^+j_{\vec q}^{sL}}\\
  = \frac{1+\varepsilon}{2}\frac TN\delta_{\vec k + \vec p - \vec q}
  (\uvec q\cdot\vec k)(S_k-1).
\end{multline}
Collecting terms one arrives at Eq.~\eqref{eq:75}.

\section{Details of the Numerics}
\label{sec:details-numerics}

For the numerical solution of
Eqs.~(\ref{eq:30},\ref{eq:80},\ref{eq:78}), we used well established
algorithms in 3D \cite{franosch+fuchs97} and 2D
\cite{bayer07}. Reciprocal space is discretized into $M^3$ grid
points ($M = 100$) up to a cutoff of $2qa = 40$ in 3D, and with $M =
125$ up a cutoff of 2qa = 50 in 2D.  The time axis is also discrete
with a grid of $N = 2048$ points and a step size that is doubled in
successive steps to accommodate for logarithmic time scales.  The
initial time step is $\Delta t = 10^{-9}t_0$. The critical density
$\varphi_c$ is located by interval bisection.

\end{document}